\documentclass[notitlepage,superscriptaddress,showpacs,nobalancelastpage,twocolumn,aps,prb,longbibliography,floatfix]{revtex4-1}%
\usepackage[english]{babel}
\usepackage{siunitx}
\usepackage[italicdiff]{physics}
\usepackage{amsfonts}
\usepackage{subfigure}
\usepackage{amsmath}
\usepackage{amssymb}
\usepackage{graphicx,epstopdf}
\usepackage{xspace}
\usepackage{array}
\usepackage[hidelinks]{hyperref}
\usepackage{color}
\usepackage{xcolor}
\usepackage{my_symbols}
\usepackage{CJK}
\usepackage{bm}
\usepackage{verbatim}
\usepackage{tikz}
\usepackage{float}
\usepackage[euler]{textgreek}
\usepackage[normalem]{ulem}
\usepackage{graphicx}%
\providecommand{\U}[1]{\protect\rule{.1in}{.1in}}

{\par}

\setcitestyle{numbers,square}
\begin{document}

\title{Circulating cavity magnon polaritons}
\author{Weichao Yu}
\email{wcyu@imr.tohoku.ac.jp}
\affiliation{Institute for Materials Research, Tohoku University, Sendai 980-8577, Japan}
\author{Tao Yu}
\email{tao.yu@mpsd.mpg.de}
\affiliation{{Max Planck Institute for the Structure and Dynamics of Matter, 22761 Hamburg, Germany}}
\affiliation{Kavli Institute of NanoScience, Delft University of Technology, 2628 CJ Delft, Netherlands}
\author{Gerrit E. W. Bauer}
\affiliation{WPI-AIMR, Tohoku University, Sendai 980-8577, Japan}
\affiliation{Institute for Materials Research, Tohoku University, Sendai 980-8577, Japan}
\affiliation{Kavli Institute of NanoScience, Delft University of Technology, 2628 CJ Delft, Netherlands}
\affiliation{Zernike Institute for Advanced Materials, Groningen University, Netherlands}
\date{\today}
\begin{abstract}
We predict magnon polariton states circulating unidirectionally in a microwave cavity when loaded by a number of magnets on special lines. Realistic finite-element numerical simulations, including dielectric, time-dependent and non-linear effects, confirm the validity of the approximations of a fully analytical input-output model. We find that a phased antenna array can focus all power into a coherent microwave beam with controlled direction and an intensity that scales with the number of magnets.
\end{abstract}

\maketitle

\section{Introduction}
The strong
magnon-photon coupling in microwave cavities
\cite{zhang_strongly_2014,huebl_high_2013,tabuchi_hybridizing_2014} allows, e.g.,  manipulation of spin currents
\cite{chumak_magnon_2015,bai_spin_2015,maier-flaig_spin_2016,bai_cavity_2017},
nonreciprocal microwave engineering \cite{wang_nonreciprocity_2019}, the design of logic devices \cite{rao_analogue_2019}, data storage \cite{zhang_magnon_2015}
and magnon entanglement for quantum information \cite{elyasi_resources_2020}.
In closed cavities the coherent coupling hybridizes magnon and photon levels into cavity-magnon--polaritons, which can be identified in terms of a level repulsion between magnon and photon levels, while a dissipative coupling in open or leaky waveguides causes level attraction
\cite{harder_level_2018,grigoryan_cavity-mediated_2019,yu_prediction_2019,xu_cavity-mediated_2019,yao_microscopic_2019}.
Analogous with structures that are coupled by optical resonators
\cite{zhang_electronically_2019}, metamaterials
\cite{baraclough_metamaterial_2019} and dielectric nanostructures
\cite{prodan_hybridization_2003,preston_vibron_2011}, multiple magnets
inside a cavity form new collective modes by the real or virtual exchange of cavity photons
\cite{zhang_magnon_2015,lambert_cavity-mediated_2016,zare_rameshti_indirect_2018}.
The polarization-momentum coupling of confined electromagnetic waves \cite{junge_strong_2013,sollner_deterministic_2015,lodahl_chiral_2017}
can be employed to realize magnet-based  broadband non-reciprocity and devices such as circulators \cite{zhu_magnon-photon_2019,zhang_broadband_2020} and a magnon accumulation in an open waveguide
\cite{yu_chiral_2019,yu_magnon_2020,yu_chiral_2020}.

\begin{figure}[th]
\includegraphics[width=0.49\textwidth]{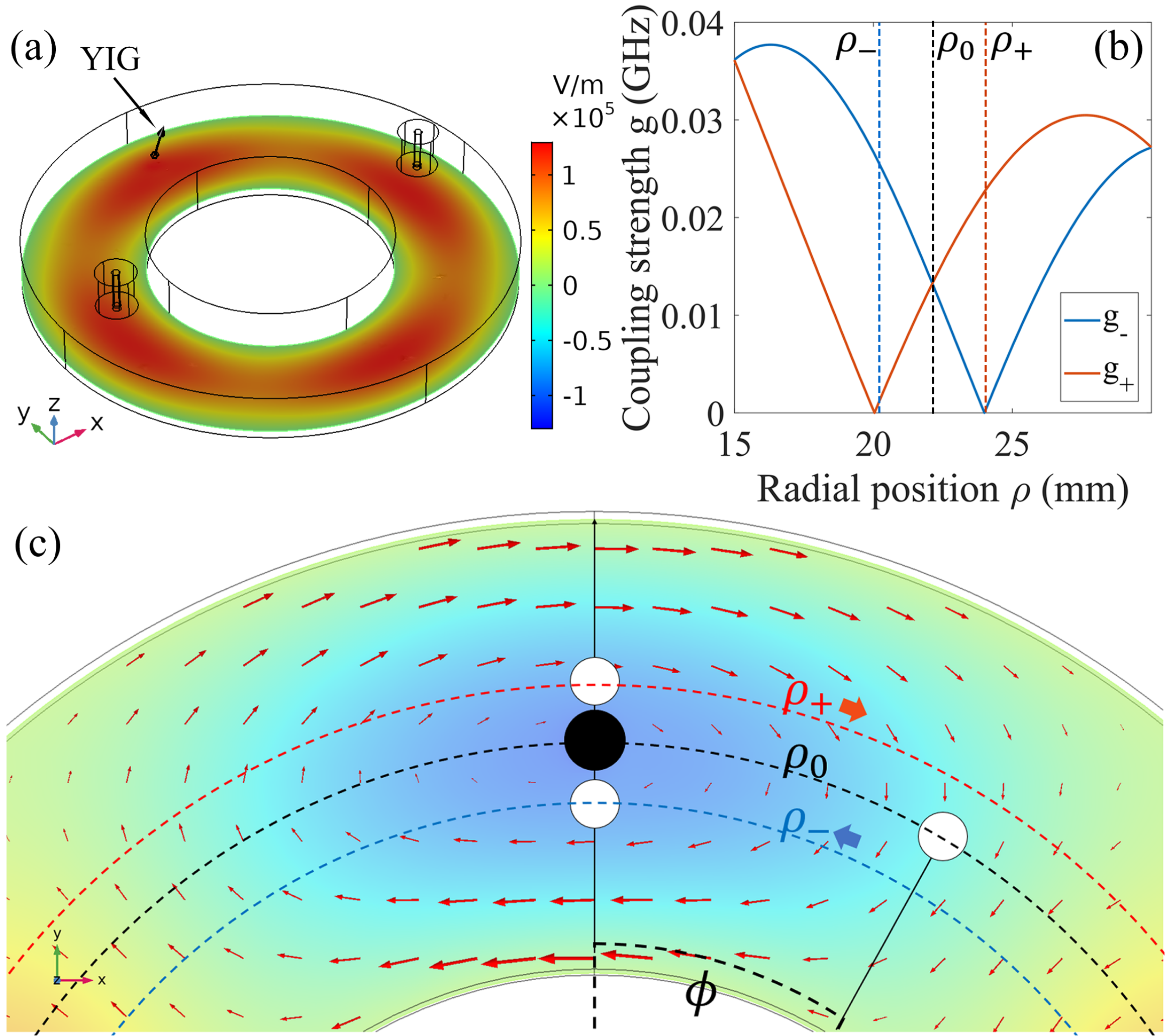}  \caption{(a) Torus-shaped
cavity ($R_{1}=15\,$mm, $R_{2}=30\,$mm, $h=6\,$mm) with two microwave ports.
A YIG sphere with diameter \SI{1}{mm} and magnetization (black arrow) saturated in
the $z$-direction is located at a distance $\rho_{+}$ from the center. The colored background encodes the computed electric field modulus $|E_{z}|$ of the $m=2$ TE cavity mode, see Appendix \ref{Torus}. (b) Magnon-photon coupling strength $|g_{\pm}|$ \Eq{coupling} as a function of the radial coordinate
$\rho$. (c) Enlarged cavity section close to the magnet, indicating the polar
angle $\phi$ and the three special positions $\rho_{0}$
and $\rho_{\pm}$ from (b). The background color is the electric field amplitude \(\Re (E_z)\)  with color code as in (a) with magnet on $\rho_{0}$, while the arrows encode the computed direction and modulus of the ac magnetic field \(\Im (\mathbf{H})\).}%
\label{fig1}%
\end{figure}

Here we explore the chiral coupling between magnets in a high-quality closed cavity, i.e., the magnon only couples to the photon circulating in one direction. We work with a torus shape illustrated in Fig.~\ref{fig1}(a) and YIG spheres with \(1 \,\)mm diameter. A chiral cavity magnon polariton state forms by putting a magnet on the special lines or plane in the cavity at which the  circular polarization of a cavity mode is locked to its propagation direction. Exciting an array of \(N\) magnets on such a line by local microwave antennae with power \(P_0\)  can generate a high-power $(\sim NP_0)$ uni-directional photon beam with high coherence and narrow band width.

In Section II we introduce the model cavity, review the basics of its dynamics without and with a magnetic load and explain the principle of the chiral coupling when magnets are put only on the special planes. We discuss the numerical solutions of the coupled Maxwell and Landau-Lifshitz Gilbert equation for a given torus cavity with up to 4 inserted magnets in Section III. These calculation justify several approximations that allow analytical calculations of, for example, the microwave scattering matrix and collective modes, as explained in Section IV. The manuscript ends with a discussion in Section V and a few Appendices with technical details.

\section{Model}
We focus on the lowest TE mode of a torus with inner
and outer radii $R_{1}$ and $R_{2}$ and height $h$ in the
coordinate system of Fig.~\ref{fig1}(a), in which the electric-field components $E_{x,y}=0$ and $E_{z}$ is homogeneous along the $\hat{\mathbf{z}}$-direction (see Appendix \ref{Torus}). The local
magnetic field  $\mathbf{H}(\rho,\phi)=H_{\rho}\hat{\mathbf{e}}_{\rho
}+H_{\phi}\hat{\mathbf{e}}_{\phi}$ reads
\begin{align}
H_{\rho}(\rho,\phi) &  =\frac{1}{\mu_{0}\gamma_{m}c}\frac{m}{\rho}%
E_{z},\nonumber\\
H_{\phi}(\rho,\phi) &  =-i\frac{1}{\mu_{0}\gamma_{m}c}\frac{\partial
E_{z}}{\partial\rho},
\end{align}
where $\mu_{0}$ and $c$ are the permeability and speed of light in vacuum,
$\{\rho,\phi\}$ are the cylindrical coordinates and $\gamma_{m}=\omega
_{m}/c$ with frequency $\omega_{m}$. The integer $m$ governs the orbital angular momentum of degenerate clockwise (CW, $m>0$)  and counter-clockwise (CCW, $m<0$ ) photon circulation. The rotation direction is ``locked" to the momentum by $\pm|m|$. At the special positions $\rho_{\pm}$ in Fig.1(b) governed by
\begin{equation}
\frac{m}{\rho_{\pm}}E_{z}(\rho_{\pm})+\left.  \frac{\partial E_{z}(\rho
)}{\partial\rho}\right\vert _{\rho=\rho_{\pm}}=0,
\end{equation}
the magnetic field is circularly polarized with $H_{\phi}=i\mathrm{sgn}%
(m)H_{\rho}$. At $\rho_0$ in Fig.1(b), $\partial_\rho E_z(\rho)\vert_{\rho=\rho_0}=0$ and the magnetic field is linearly polarized. \Figure{fig1}(c) shows the  $m=2$ TE mode, in which the colored background represents the computed \(\Re (E_z)\) and the arrows the direction and magnitude of \(\Im (\mathbf{H})\).

We  load the cavity with $N$ magnetic spheres centered at $(\rho,\phi_{l})$ ($l\in \{1,\cdots, N\}
$).  The magnetization $\mathbf{M}_{l}(\mathbf{r})$ couples to the
microwaves through the Zeeman interaction. The magnets are saturated to
$M_{s}\hat{\mathbf{z}}$ by a static field $H_0=0.3895\,$T. The transverse dynamics $\mathbf{m}$ of a magnet on \(\rho_{\pm}\), couples only to one of the counter rotating cavity modes. The magnets interact with each other via the cavity modes and form collective states that can be selectively excited by an array of local antennas attached to the magnetic spheres.

We compute the dynamical properties both in the frequency and time domain (see Appendix \ref{Numerics}). In the former, we numerically solve the coupled Maxwell and linearized Landau-Lifshitz-Gilbert (LLG) equation in
the frequency domain in the macrospin and rotating wave approximations \cite{cao_exchange_2015,zare_rameshti_indirect_2018}
\begin{equation}
i\omega\mathbf{m}=\hat{\mathbf{z}}\times(\omega_{M}\mathbf{H}-\omega
_{\mathrm{K}}\mathbf{m}+i\alpha\omega\mathbf{m}),\label{eqn:1}%
\end{equation}
where $\omega$ is the angular frequency, $\omega_{M}=\gamma M_{s}$ with $\gamma$ the (modulus) of the
gyromagnetic ratio, $\alpha$ is the Gilbert damping constant, and
$\omega_{\mathrm{K}}=\gamma H_{0}$ is the Kittel mode frequency which linearly
depends on the external field $H_{0}$ \cite{kittel_theory_1948}. The solution
of \Eq{eqn:1}, $\mathbf{m}=\mu_0\overline{\zeta}\mathbf{H}$, defines the
susceptibility \(\overline{\zeta}\) and the permeability (the overlines denote tensors)
\begin{equation}
\overline{\mu}_\text{M}=\overline{\text{I}}+\overline{\zeta}=\left(
\begin{array}
[c]{ccc}%
1+u & -iv & 0\\
iv & 1+u & 0\\
0 & 0 & 1
\end{array}
\right),
\end{equation}
where $u={\left(
\omega_{\mathrm{K}}-i\alpha\omega\right)  \omega_{M}}/[{\left(  \omega
_{\mathrm{K}}-i\alpha\omega\right)  ^{2}-\omega^{2}}]$ and $v={\omega
\omega_{M}}/{[\left(  \omega_{\mathrm{K}}-i\alpha\omega\right)  ^{2}%
-\omega^{2}]}$ and the driving magnetic field $\mathbf{H}$ is injected by the ports and generated at local antennae.  The presence of the magnetic spheres
affects the microwave by the spatially dependent relative
permittivity $\varepsilon_{r}(\mathbf{r})$ and permeability $\mu
_{r}(\mathbf{r})$ through Maxwell's equation \cite{jackson_classical_1998},
e.g.,
\begin{equation}
\nabla\times\lbrack\mu_{r}(\mathbf{r})^{-1}\nabla\times\mathbf{E}%
]-k^{2}\varepsilon_{r}(\mathbf{r})\mathbf{E}=0,\label{eqn:4}%
\end{equation}
where $k=\omega/c$ is the wave number of light in vacuum. For YIG $(\varepsilon_r,\mu_r)=(15,\overline{\mu}_{\text{M}})$ \cite{sadhana_synthesis_2009} inside the magnets and $(\varepsilon_r,\mu_r)=(1,1)$ in the rest of the cavity.

The energy transported by
propagating electromagnetic waves is captured by the cycle-averaged Poynting vector
$\mathcal{P}=\frac{1}{2}\text{Re}\left(  \mathbf{E}^{\ast}\times
\mathbf{H}\right)  $ \cite{jackson_classical_1998},
where the asterisk symbol denotes the complex conjugate. The Poynting vector encodes both the direction and modulus of the energy flow and is proportional to the linear momentum density with $\mathbf{p}=\mathcal{P}/c^2$. The latter can be separated into an orbital and a spin contribution $\mathbf{p}=\mathbf{p}_{\text{o}}+\mathbf{p}_{\text{s}}$ \cite{berry_optical_2009,aiello_transverse_2015,aiello_ubiquitous_2016,bliokh_optical_2017,bliokh_transverse_2015}.  The orbital momentum $\mathbf{p}_0$ reads \begin{equation}
\mathbf{p}_{\text{o}}=\frac{1}{4\omega}\text{Im}\left[  \varepsilon
_{0}\mathbf{E}^{\ast}\cdot\nabla\mathbf{E}+\mu_{0}\mathbf{H}^{\ast}%
\cdot\nabla\mathbf{H}\right],  \label{CanonicalMomentum}
\end{equation}
where $\varepsilon_{0}$ and $\mu_{0}$ are the vacuum permittivity and permeability. The spin part of the linear momentum density $\mathbf{p}_{\text{s}}=\frac{1}
{2}\nabla\times\mathbf{s}$,
where $\mathbf{s}$ is the spin angular momentum (SAM) density,
\begin{equation}
\mathbf{s}=\frac{1}{4\omega}\text{Im}\left[  \varepsilon_{0}\mathbf{E}^{\ast
}\times\mathbf{E}+\mu_{0}\mathbf{H}^{\ast}\times\mathbf{H}\right]
\label{SpinAngularMomentum}%
\end{equation}
which at GHz frequencies is dominated by the magnetic field component. A finite SAM implies a photonic energy and momentum flow.

\section{Numerical results}

The CW and CCW TE modes form standing wave modes by the perturbation formed by the ports or the magnet. This normal scattering competes with the chiral coupling between magnetic field and the dynamic magnetization. Both contribute to the mixing with other cavity modes which are included in the numerical calculations. Nevertheless, in the present configuration the $m=\pm2$  cavity modes with frequency $\omega_c=\omega_{m=\pm2}=$\SI{10.84}{GHz} dominate.


\begin{figure*}[tbh]
\caption{(a)-(d), Transmission power spectrum $\vert S_{21} \vert ^{2}$ calculated numerically for a single magnetic sphere located in the cavity at (a) $\rho_-$, $\phi$=0, (b) $\rho_+$, $\phi$=0, (c) $\rho_0$, $\phi$=0 and (d) $\rho_0$, $\phi$=$\pi/4$. The dashed
vertical line is at $H_0=\SI{0.3895}\rm{T}$. (e) Integrated spin angular momentum (SAM) \Eq{SpinAngularMomentum}) across the cavity cross section for a magnet at  $\rho_{-}$, $\rho_{0}$ and $\rho_{+}$ as a function of microwave input frequency. Insets: SAM density polarized along \textbf{z} and orbital momentum density (black arrows indicate direction and modulus) \Eq{CanonicalMomentum} in the cavity plane. Letters A-F label the resonance peaks in \Figure{fig2}(a-b).  }%
\label{fig2}%
\centering
\includegraphics[width=0.99\textwidth]{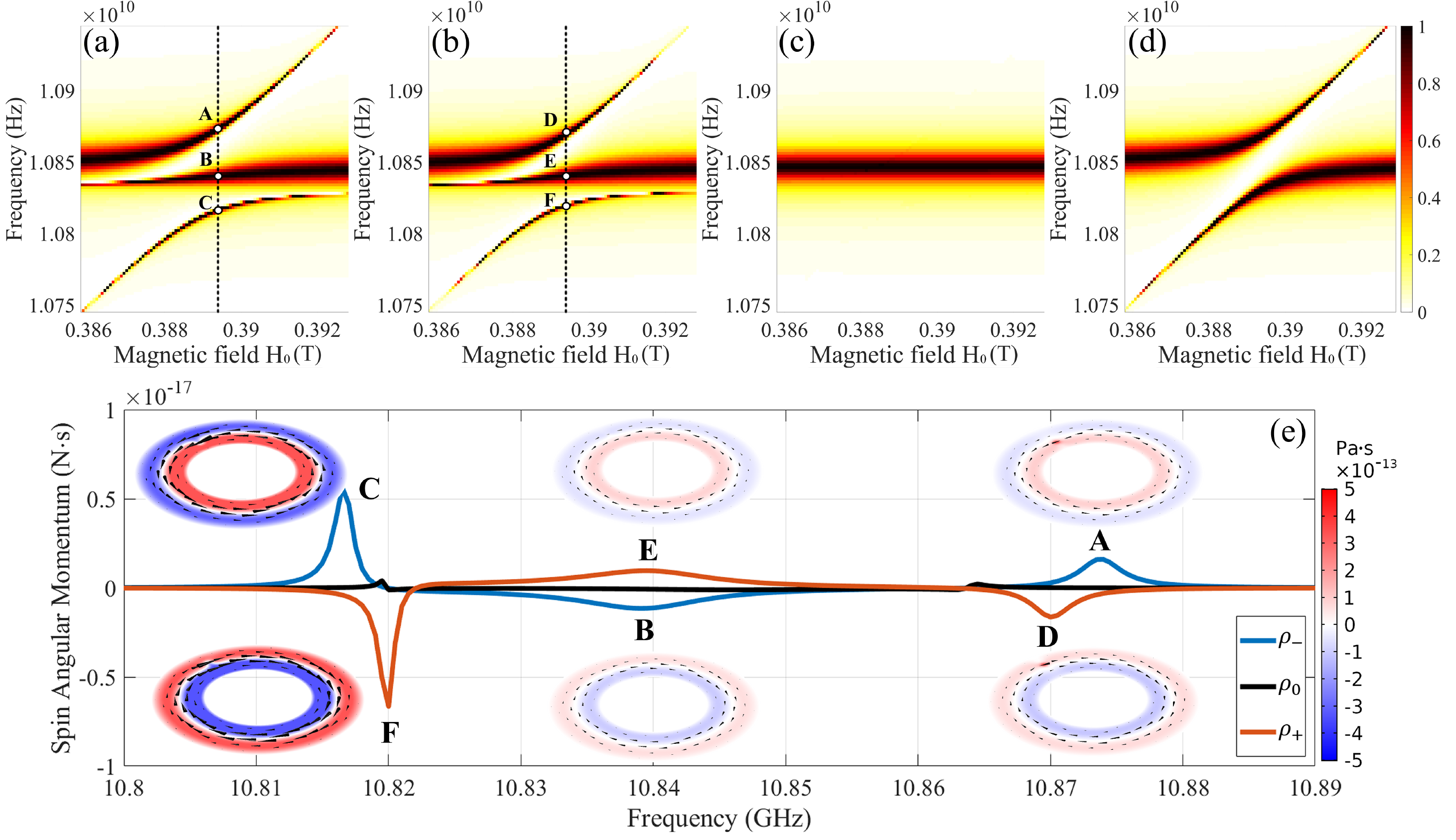}
\end{figure*}
The calculated microwave transmission spectra \(S_{21}(\omega)\)  (see Appendix \ref{Numerics}) through the ports 1 and 2 in \Figure{fig2}(a)-(d) for a cavity loaded with a single sphere contain detailed information about the magnon polariton. In \Figure{fig2}(a) the macrospin (Kittel) mode in a magnet at  $\rho_{-}$ couples strongly with the CCW mode with a splitting indicated by (A, C), while the CW mode appears to not interact (B). The magnet at $\rho_{+}$ only couples to the CW mode (\Figure{fig2}(b)). The gap is smaller than in \Figure{fig2}(a), consistent with a larger circumference of the circle with radius $\rho_{+}$. The spectra do not depend visibly on the polar position \(\phi\) of the magnet (not plotted here).  The double peak structure far from the anticrossing is caused by the normal scattering between the \(m=\pm2\) modes. A magnet on $\rho_{0}$ interacts with both propagating CW and CCW modes but now the coupling depends on \(\phi\), another evidence for normal scattering that pins a standing wave magnetic field distribution \(\sim \sin 2 \phi\) with a maximum at \(\phi=\pi/2\) and a node at  \(\phi=0\) as observed in \Figure{fig2}(c)(d) and also as a modulation of the electric field in \Figure{fig1}(a). These effects are relatively small because the size of the magnet (\SI{1}{mm}) is much smaller than the wavelength divided by the dielectric constant of the sphere $\lambda/\epsilon\sim\SI{9.4}{mm})$.  The magnetic sphere can be treated as a point particle, while the chiral coupling overwhelms the normal scattering when the magnets are on special lines $\rho_{\pm}$, which allows to adapt below the analytic treatment introduced by Yu \textit{et al.} \cite{yu_chiral_2019,yu_magnon_2020,yu_chiral_2020}  for a straight wave guide.

In our configuration the SAM ( \Eq{SpinAngularMomentum}) is transverse, i.e. perpendicular to the wave propagation.  Physically, the SAM is the local degree of microwave circular polarization which has a node at \(\rho_0\) and extrema at  \(\rho_\pm\), as plotted in the insets of \Figure{fig2}(e) for the resonances labeled A-F in \Figure{fig2}(a)(b). Since \(\dot{\mathbf{s}}=0\), the photon spin current is conserved. The finite curvature shifts the resonance frequencies for C and F as well as A and D and the contribution from the outer region wins in the integral over the cross section  in \Figure{fig2}(e) and peaks at the resonances. For a magnet on $\rho_{-}$ (Fig.2(a)), the hybridized modes A and C propagate CCW but when on $\rho_{+}$, F and D move CW.
We observe finite magnet size effects, viz. (i) the nominally uncoupled modes B and E acquire a weak chirality opposite to that of the strongly coupled modes, (ii) signals when the magnet is on $\rho_{0}$.

\begin{figure}[t]
\centering
\centerline{\includegraphics[width=9.3cm]{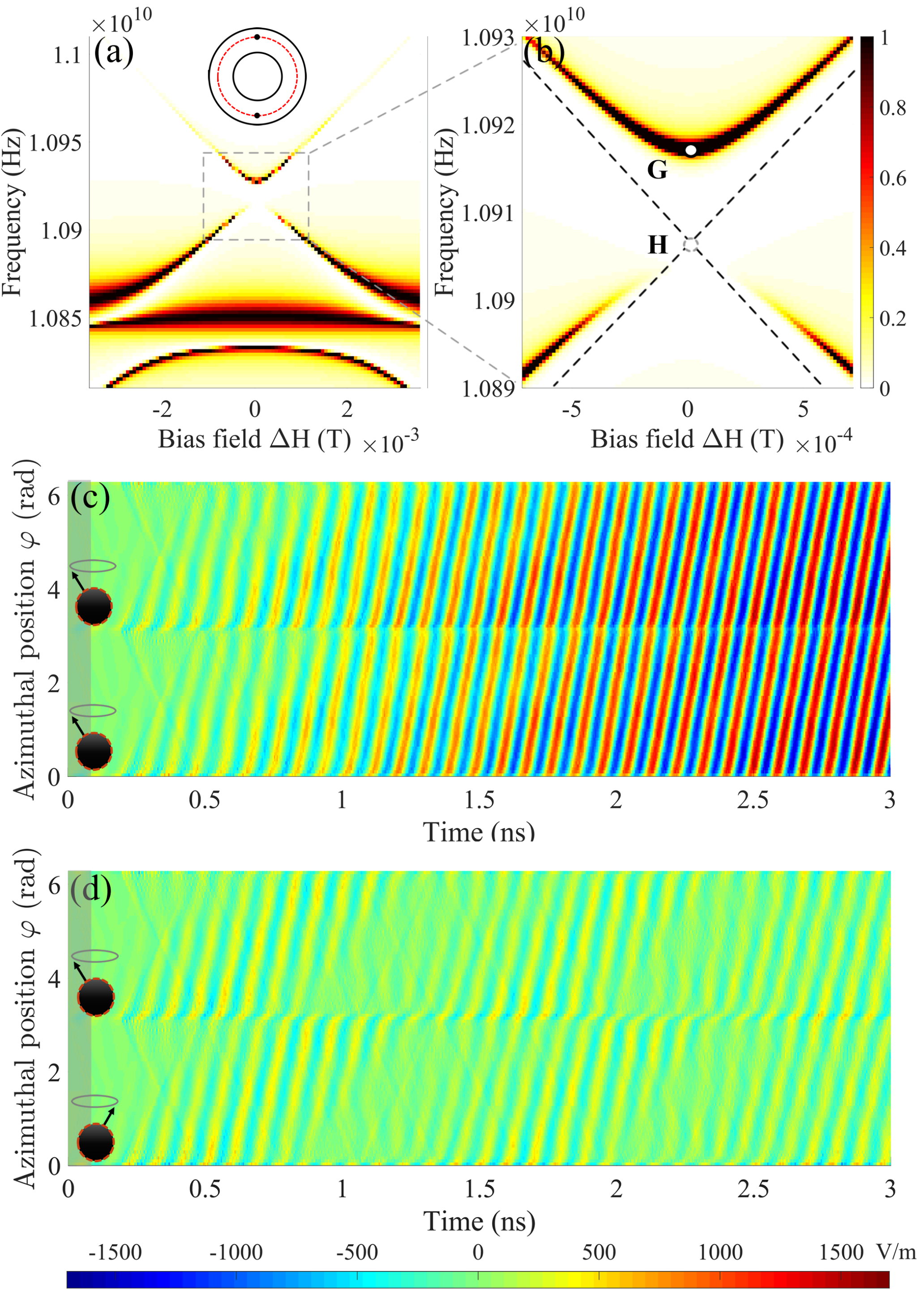}}\caption{(a) Transmission
power spectrum \(|S_{21}|^{2}\) of a magnonic dimer with applied field
$H_0=$\SI{0.3913}{T} when both spheres are on $\rho_{+}$ with \(\phi=0,\pi \). (b) Enlarged view of the anti-crossing levels. (c)-(d) Simulation of the time-dependent dynamics of the cavity photons ($E_{z}$ component)
emitted by the magnonic dimer. In (c) the bright mode (the G spot in (b)) is excited by a local transverse magnetic field pulse (duration indicated by grey shading) with the same phase on both magnets, while in (d) an out-of-phase pulse excites the dark mode (H point in (b)).}
\label{fig3}%
\end{figure}

Next, we consider the magnonic dimer, i.e. two spheres ($N=2$) at $\rho_{+}$ on opposite sides of the torus with an FMR at \SI{10.904}{GHz} detuned from  the $m=2$ cavity mode (\SI{10.84}{GHz}). At this distance, the  direct magnetodipolar interaction between the magnets is negligibly small. A local bias field with opposite sign on each sphere $H_0\pm\Delta H$ breaks the symmetry and mixes the bright acoustic (G) and dark optical modes (H) \cite{zare_rameshti_indirect_2018} when \((\Delta H \ne 0)\).


We compute the non-stationary dynamics by solving the coupled Maxwell and Landau-Lifshitz-Gilbert equations in the time domain (see Appendix\ref{Numerics}). Figure \ref{fig3}(c) and (d) show the spatio-temporal propagation of the microwave \(E_z\)-field along \(\phi\) on the circle $(\rho_{0},h/2)$ for the set-up of \Figure{fig3}(a). We excite the magnets off-resonantly by a transverse magnetic field pulse of the form $\sin(2\pi f_0 t)$, with frequency $f_0$  close to the FMR \SI{10.9}{GHz} and duration 1/$f_0$. Equal phases on both magnets excited the bright state (G spot in \Figure{fig3}(b)). The magnetization ``pumps" cavity photons that accumulate over an RC time constant ($\sim$\SI{1}{ns}) that is governed by the ratio of the coupling strength and dissipation rate. The (nearly) steady state decays on a ms scale governed by the weak damping. The constant slope  \(\dot{\phi} \approx 2 \pi / (100 \, \rm{ps}) = 2\pi \times 10 \, \rm{GHz} \) of the phase patterns in \Figure{fig3}(c)) is consistent with the photon phase velocity  $v_{\text{phase}}=7.5\times10^8\,$m/s, where $v_{\text{phase}}=\omega / \sqrt{k^2-k_c^2}$, $k=\omega / v_\text{vacuum}$, $k_c= \pi /a $, the width of waveguide $a=15\,$mm, and $\omega=2\pi\cdot10.918\,$GHz.  $v_{\text{phase}}>c$, but the group velocity of energy transport $v_\text{group}=c^2/v_\text{phase}=1.2\times10^8\,$m/s. Dark mode magnons excited by out of phase local fields (H point in \Figure{fig3}(b)) and \Figure{fig3}(d) hardly generate any cavity photons, as expected.

\begin{figure}[tbh]
\includegraphics[width=9.2cm]{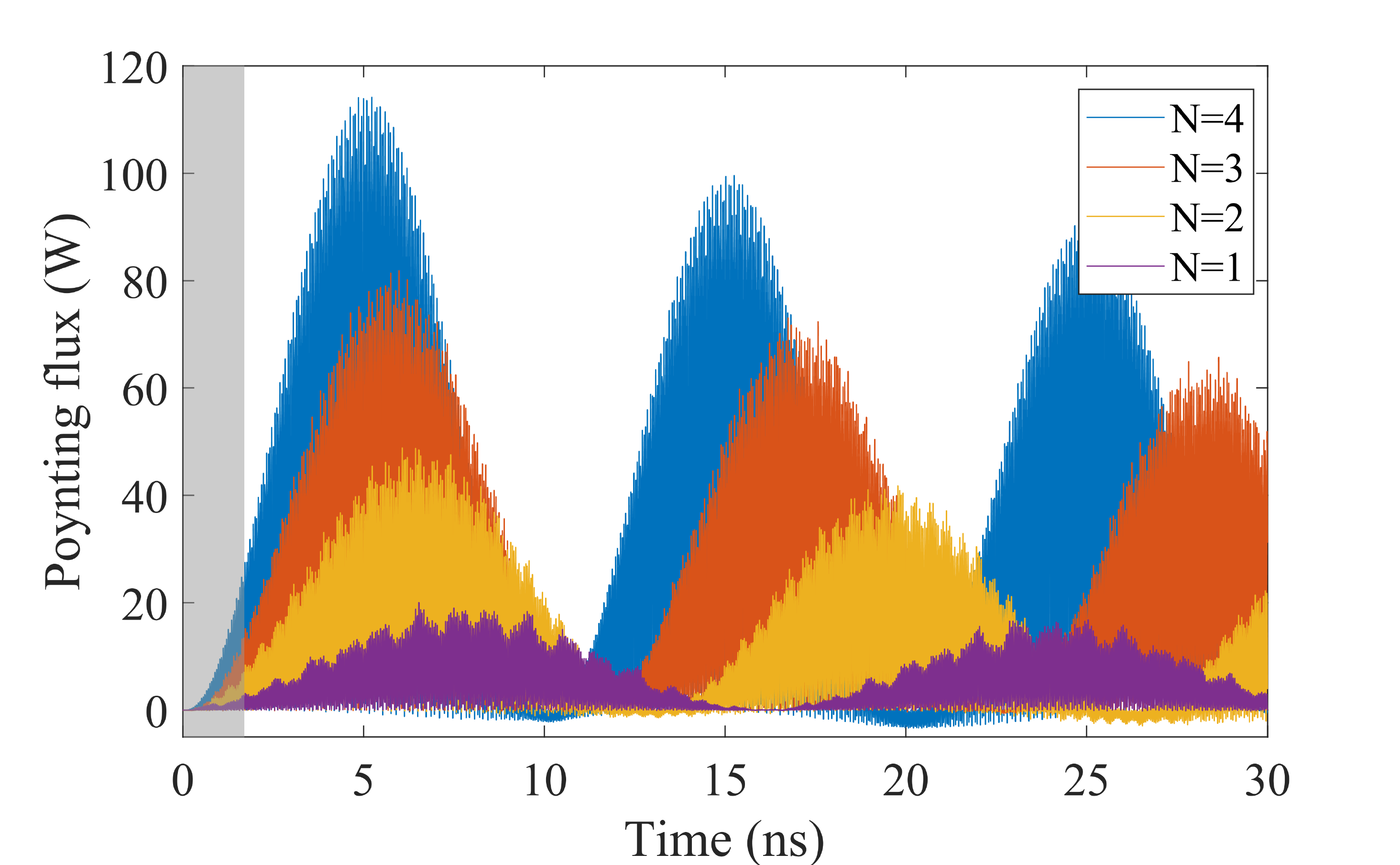}\caption{Time-dependent Poynting flux $P_\text{flux}$ calculated for 1 to 4 magnets equidistantly located on the circle $\rho_+$. The local pulses on each sphere have the form $\mathbf{h}_\text{local}=(A\cos(\omega t+\varphi_l),A\sin(\omega t+\varphi_l),0)$ with amplitude $A=$\SI{1000}{A/m}, frequency $\omega/2\pi=10.82\,$GHz and phase $\varphi_l=6(l-1)\pi/N$ (duration $T=2/f_0=0.18 \,\rm{ns}$ as indicated by the grey shading). \label{fig:PulsePower}}%
\end{figure}

Figure \ref{fig:PulsePower} shows results from time-dependent simulations for different numbers of magnets. The Poynting flux, i.e. the integral of the Poynting vector $P_\text{flux}=\mathbf{n}\cdot\int_\Omega \mathcal{P}d\rho dz=\mathbf{n}\cdot\int_\Omega  \mathbf{E}\times\mathbf{H}d\rho dz$  over the cavity cross section $\Omega$, is the circulating power and \(P_\text{flux}>0\) in the CW direction. Here we drive $N\in \{1,2,3,4\}$ magnets by circularly polarized pulses from local antennas with phases matched to the bright magnon polariton state. The local fields excite only magnons (not magnon polaritons) so we observe Rabi beatings between magnons and photons. The maximum microwave power and Rabi frequency scale linearly with $N$.

\begin{figure}[tbh]
\includegraphics[width=9.2cm]{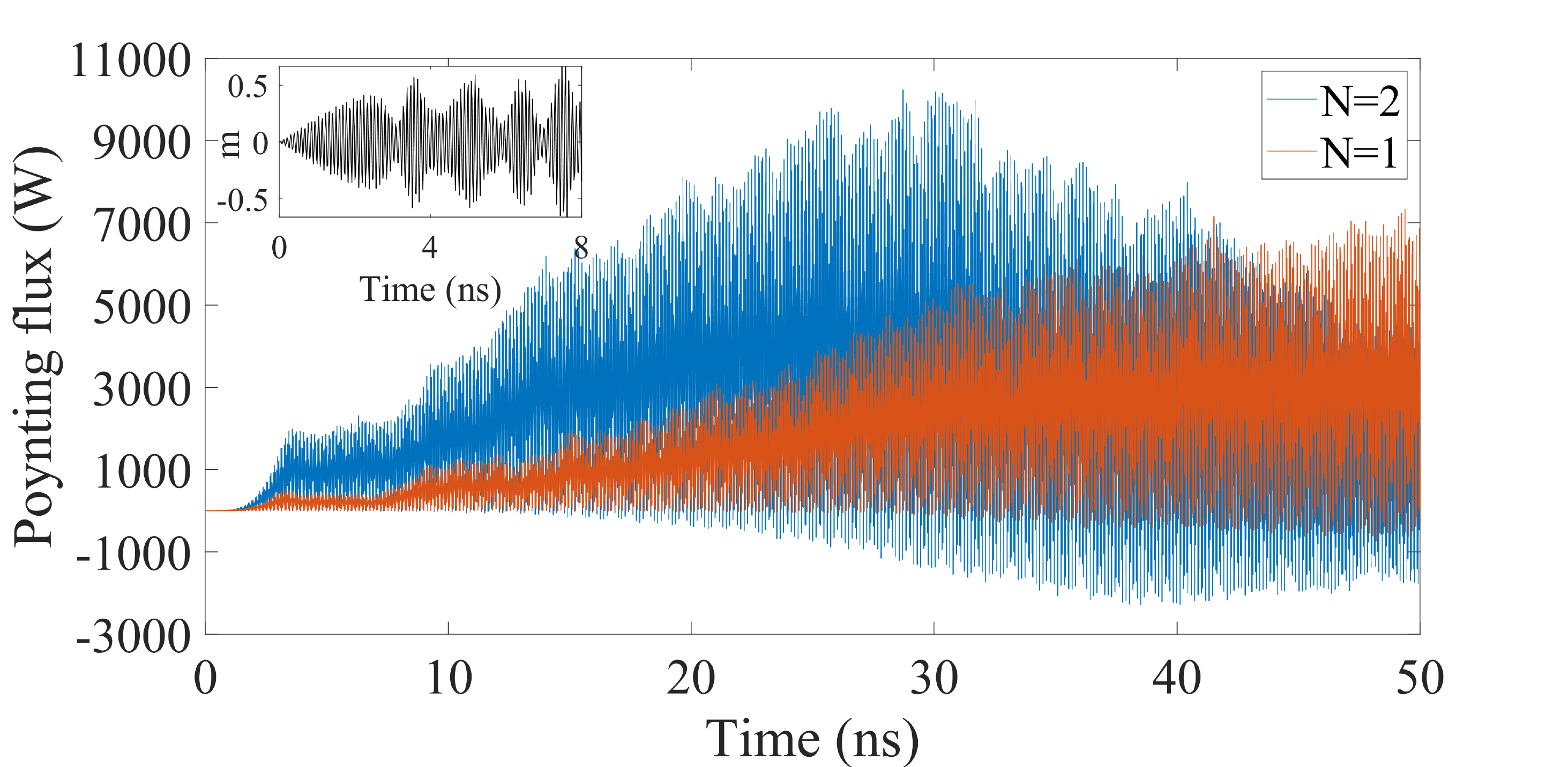}
\caption{Nonlinear dynamics of strongly driven magnon polaritons. The Poynting flux $P_\text{flux}$ is plotted in red for $N=1$ and blue for $N=2$. The magnetic sphere on $\rho_+$ are continuously driven by a circularly polarized local field $\mathbf{h}_{\text{local}}(t)=(A\cos\omega t,A\sin\omega t,0))$ with magnitude $A=$\SI{1000}{A/m} and frequency $\omega/2\pi=$\SI{10.82}{GHz} that is switched on at \(t=0\). For $N=2$ case the  bright state is excited. Inset: in-plane component of the precession for $N=1$.
\label{fig:nonlinear}}%
\end{figure}

When under high excitation the in-plane magnetization component reaches $\sim0.4$,  (precession cone angle $\sim24\rm ^{\circ}$), the linear approximation breaks down and the power output does not linearly scale with the number of spheres $N$ anymore. We can access these effects since we solve the dynamics of the magnon-photon hybrid system without linearization or rotating wave approximation, keeping in mind that we are still in the macrospin approximation. In \Figure{fig:nonlinear} we drive a single magnetic sphere by a sustaining local source that is switched on at \(t=0\). The nonlinear regime is reached after about $t=3\,$ns. We estimate the maximum power, by integrating the Poynting flux over the cross section.
A single magnet can pump \SI{500}{W} in the linear and \SI{7000}{W} in the chaotic regime (red signal in \Figure{fig:nonlinear}) into a unidirectional mode. Two spheres inject \SI{1300}{W}  in the linear and up to \SI{10000}{W} in the chaotic regime (blue signal in \Figure{fig:nonlinear}). The spheres are on $\rho_+$ and nominally couple only to the CW mode. The negative Poynting flux indicates CCW mode excitation in the chaotic regime.

\section{Analytical input-output model}

The simulations show that the Hilbert space of the problem is to a good approximation limited to two cavity modes, magnets behave like point particles, and cavity mode mixing is very small. This allows us to build a simple yet accurate model for the collective modes with emphasis on the power output for an arbitrary number of magnetic inserts. We use a quantum description for convenience, noting that classical picture is recovered by replacing operators by field amplitudes. The following holds in the linear regime and we adopt the rotating wave approximation that is valid outside the ultrastrong coupling regime. We refer for technical details to Appendix \ref{QPhoton}.

The system Hamiltonian $\hat{H}=\hat{H}_{p}+\hat{H}_{m}+\hat{H}_{\mathrm{int}}$ has contributions from the photons, magnons, and their interaction, respectively. The cavity photons are harmonic oscillators with frequency \(\omega_{m}\) and creation/annihilation operators \(\hat{\alpha}_{m}^{\dagger}/\hat{\alpha}_{m}\) (Appendix \ref{QPhoton})
 \begin{equation}
\hat{H}_{p}=\sum_{m}\hbar
\omega_{m}\hat{\alpha}_{m}^{\dagger}\hat{\alpha}_{m}.\label{Photon}%
\end{equation}At not too high excitations, \(N\) identical magnets can be modelled as a sum of degenerate harmonic oscillators with Kittel mode frequency \(\omega_{\mathrm{K}}\) and creation/annihilation operators \(\hat{m}_{l}^{\dagger}/\hat{m}_{l}\) for the \(l\)-th magnet:
\begin{equation}
\hat{H}_{m}=\hbar\omega_{\mathrm{K}}\sum_{l=1}^{N}\hat{m}%
_{l}^{\dagger}\hat{m}_{l}.\label{Magnon}%
\end{equation}The Zeeman interaction between the photon magnetic field and the magnetic inserts reads \begin{equation}
\hat{H}_{\mathrm{int}}=\hbar\sum_{m}\sum_{l=1}^{N}|g_{l,m}|e^{im_\ast\phi_l}\hat{\alpha}_{m}%
\hat{m}_{l}^{\dagger}+\mathrm{H.c.}\label{MagnonPhotonCoupling}%
\end{equation}with \(m_\ast=m+1\), and coupling strengths
\(g_{l,m}\) \cite{yu_magnon_2020} (refer to Appendix \ref{MM}) that depend on position $\rho_l$ and magnitude as well as
polarization of the local field, as shown in \Figure{fig1}(b). The net increase of 1 between $m_\ast$ and $m$ reflects the circulating topology of photon states. We do not restrict the equal spacing of the magnets, i.e., an abitrary distribution of $\phi_l$. On the
special circles $\rho_{+}$ ($\rho_{-}$), one of the coupling strength $|g_{-}(\rho_{+})|$ ($|g_{+}(\rho_{-})|$) vanishes. The pure chiral coupling $|g_{-}(\rho_{-})|=0.026\,\mathrm{{GHz}}$ on $\rho_{-}$ and
$|g_{+}(\rho_{+})|=0.023\,\mathrm{{GHz}}$ on $\rho_{+}$ calculated for 1 mm YIG spheres and microwave amplitudes of the empty cavity agrees with the level splittings in Fig.~2(a) and (b).

The dynamics of the conservative system is governed by the Heisenberg equation of motion corresponding to the model Hamiltonian augmented by sources (drive) and sinks (dissipation) terms. When all magnets are on \(\rho_+\) the right circulating photon $\hat{\alpha}_{+}$ couples to both the cavity
input \(\hat{Q}_{\mathrm{in}}=\sqrt{\kappa_r}\hat{q}_\mathrm{in}\) via the coupling rate \(\kappa_r\)  and the magnons. We may disregard the other mode $\hat{\alpha}_{-}$ since we only focus on the chiral coupling to the cavity mode \(m=+2\). The equation of motion reads
\begin{align}
i\frac{d\hat{\alpha}_{+}}{dt} &  =\omega_{m}\hat{\alpha}_{+}-i\frac
{\tilde{\kappa}_{\alpha}}{2}\hat{\alpha}_{+}+|g_{+}|\sum_{l}e^{-im_{\ast}%
\phi_{l}}\hat{m}_{l}-i\hat{Q}_{\mathrm{in}},\nonumber\\
i\frac{d\hat{m}_{l}}{dt} &  =\omega_{\mathrm{K}}\hat{m}_{l}+|g_{+}%
|e^{-im_{\ast}\phi_{l}}\hat{\alpha}_{+}-i\frac{\tilde{\kappa}_{m}}{2}\hat
{m}_{l}-i\hat{P}_{l},\label{EOM_mp}%
\end{align}
where $\tilde{\kappa}_{\alpha}=\kappa
_{\alpha}+\kappa_{r}$ and  $\tilde{\kappa}_{m}=\kappa
_{m}+\delta_{r}$ describe dissipation in the form of the sum of intrinsic and radiative damping of photons and magnons, respectively, and $\hat{P}_{l}=\sqrt{\delta_{r}}\hat{p}_{l}$ with local antenna input $\hat{p}_{l}$. The output amplitude at the port  \(\hat{p}_{\mathrm{out}}=\sqrt
{\kappa_{r}}\hat{\alpha}_{+}\) is governed by the same radiative damping $\kappa_{r}$ as the input one since we consider identical ports in this work. To solve the equation of motion, we use the Bloch ansatz $\hat{\Psi}_{\chi}%
(\omega)=\sum_{l=1}^{N}e^{i{\chi}\phi_{l}}\hat{m}_{l}%
(\omega)/\sqrt{N}$ with ``crystal momentum'' $\chi$, leading to
\begin{align}
&  \left(  \omega-\omega_{\mathrm{K}}+i\frac{\tilde{\kappa}_{m}}{2}\right)
\hat{\Psi}_{\chi}-\Gamma_{+}(\omega)A_{m_{\ast}}(\chi)\hat{\Psi}_{m_{\ast}%
}\nonumber\\
&  =-\frac{i}{\sqrt{N}}A_{m_{\ast}}(\chi)\frac{\Gamma_{+}(\omega)}{|g_{+}%
|}\hat{Q}_{\mathrm{in}}(\omega)-\frac{i}{\sqrt{N}}\sum_{l}e^{i\chi\phi_{l}%
}\hat{P}_{l}(\omega).
\end{align}
As demonstrated later, when the magnets at \(\phi_l\) are equally spaced, $\chi\in\mathbb{Z}_{0}$ by the translation symmetry; otherwise, $\chi$ is not generally an integer.
Here $\Gamma_{+}(\omega)={|g_{+}|^{2}}/\left(  {\omega
-\omega_{m}+i\tilde{\kappa}_{\alpha}/2}\right)$ is the photon-mediated effective magnon-magnon interaction and $A_{m}(\chi)\equiv\sum_{l=1}^{N}e^{i(\chi-m_{\ast})\phi_{l}}$ is a cumulative phase factor.

 We see that $\chi=m_{\ast}$ is a solution with $A_{m_{\ast}}(m_{\ast})=N$ and $m_{\ast}$  labels the ``bright" cavity magnon polariton whose frequencies obey
\begin{equation}
\left(  \omega-\omega_{\mathrm{K}}+i\frac{\tilde{\kappa}_{m}}{2}\right)
\left(  \omega-\omega_{m}+i\frac{\tilde{\kappa}_{\alpha}}{2}\right)
=N|g_{+}|^{2},\label{MP}%
\end{equation}
with a coupling strength enhanced by $\sqrt{N}$ compared to a single magnet but still with a damping of one magnet. $\chi=m_\ast$ and the associated frequency are solutions even when the polar coordinate $\phi_l$  are randomly distributed. All other solutions are degenerate at  $\omega_{\mathrm{K}}-i\tilde{\kappa}_{m}/2$ since $A_{m_{\ast}}(\chi \ne m_{\ast})=0$, enforcing that
\begin{align}
\sum_{l=1}^{N}e^{i(\chi-m_{\ast}%
	)\phi_{l}}=0
\end{align}
and the roots determine the ``momentum" of the other modes.
 They are ``dark'', i.e., do not couple to the cavity photons. The degeneracy implies a flatband with respect to the ``momentum". These solutions are readily applied to the quantum emitters such as atom electric dipoles in a closed cavity as well. In the dispersive regime with a detunning between the magnon and photon modes, the photon can mediate an effective coupling between magnons, i.e., $\Gamma$ in the above. The dispersive energy gap between bright and dark modes in \Figure{fig3} can be estimated by $N\Gamma_{+}=N|g_{+}|^{2}(\omega
_{c}-\omega_{\mathrm K})/\left[  (\omega_{c}-\omega_{\mathrm K})^{2}+(\kappa/2)^{2}\right]
=\SI{0.013} \, \rm{GHz}$ for a joint magnon and photon dissipation $\kappa/2=\SI{3.88}\,\rm{MHz}$ and $\omega_c=$\SI{10.84}{GHz}, agree well with the numerical experiments.

For a general excitation by the ports and local sources, the excited magnon amplitudes
$\mathcal{M}=%
\begin{pmatrix}
\hat{m}_{1}(\omega),\hat{m}_{2}(\omega),\cdots,\hat{m}_{N}(\omega)
\end{pmatrix}
^{T}$ is
\begin{equation}
\mathcal{M}=\Psi_{m_{\ast}}(\omega)%
\begin{pmatrix}
e^{-im_{\ast}\phi_{1}},e^{-im_{\ast}\phi_{2}},\cdots,e^{-im_{\ast}\phi_{N}}%
\end{pmatrix}
^{T},
\end{equation}
with bright mode magnon amplitude
\begin{equation}
\hat{\Psi}_{m_{\ast}}(\omega)=\frac{-i\sqrt{N}\frac{\Gamma_{+}(\omega)}%
{|g_{+}|}\hat{Q}_{\mathrm{in}}(\omega)-i\frac{1}{\sqrt{N}}\sum_{j}e^{im_{\ast
}\phi_{j}}\hat{P}_{j}(\omega)}{\omega-\omega_{\mathrm{K}}+i{\tilde{\kappa}%
_{m}}/{2}-N\Gamma_{+}(\omega)}.
\end{equation}
and cavity photon amplitude
\begin{equation}
\hat{\alpha}_{+}(\omega)=\frac{|g_{+}|\sqrt{N}\hat{\Psi}_{m_{\ast}}%
(\omega)-i\hat{Q}_{\mathrm{in}}(\omega)}{\omega-\omega_{m}+i\tilde{\kappa
}_{\alpha}/2}.\label{alpha_R}%
\end{equation}
With only global input (\(\hat{Q}_{\mathrm{in}}\ne0\), $\hat{P}_{j}=0$), this mode is split-off from the degenerate magnon and photon modes by the resonant polariton gap $\sqrt{N}|g_{+}|$.

The microwave
transmission between the ports communicated by CW modes reads
\begin{equation}
S_{12}(\omega)=\frac{-i\kappa_{r}}{\omega-\omega_{m}+i\frac{\tilde{\kappa
}_{\alpha}}{2}-\frac{N|g_{+}|^{2}}{\omega-\omega_{\mathrm{K}}+i{\tilde{\kappa
}_{m}}/{2}}},
\end{equation}
Of special interest is the output power $|\hat{p}_{\mathrm{out}%
}|^{2}={\kappa}_{r}|\hat{\alpha}_{+}|^{2}$ when microwaves are injected via the local antenna array (and $\hat{Q}_{\mathrm{in}}=0$) phase matched to the bright mode as $\hat{P}_{j}(\omega)=\hat{P}_{0}(\omega)e^{-im_{\ast}\phi_{j}}$. The amplitude for either one of the two magnon polariton frequencies $\omega_{\mathrm{P}}=(\omega
_{m}+\omega_{\mathrm{K}})/{2}\pm\sqrt{\left(  {\omega_{m}-\omega_{\mathrm{K}}%
}\right)  ^{2}/4+N|g_{+}|^{2}}$  reads
\begin{equation}
\hat{\alpha}_{+}(\omega_{\mathrm{P}})=\frac{-{2}N|g_{+}|\hat{P}_{0}%
(\omega_{\mathrm{P}})}{\tilde{\kappa}_{\alpha}(\omega_{\mathrm{P}}%
-\omega_{\mathrm{K}})+{\tilde{\kappa}_{m}}(\omega_{\mathrm{P}}-\omega
_{m})+i\tilde{\kappa}_{\alpha}{\tilde{\kappa}_{m}}/{2}}%
\end{equation}
For $\tilde{\kappa}_{\alpha}\gg\tilde{\kappa}_{m}$ and $\omega_{m}%
=\omega_{\mathrm{K}}$, and disregarding out-of-phase term,
we arrive at
\begin{equation}
\hat{\alpha}_{+}(\omega_{\mathrm{P}})\rightarrow\mp\frac{2\sqrt{N}\hat{P}%
_{0}(\omega_{\mathrm{P}})}{\tilde{\kappa}_{\alpha}}+O(\kappa^{2}).
\end{equation}
The output power $|\hat{p}_{\mathrm{out}%
}|^{2}={\kappa}_{r}|\hat{\alpha}_{+}|^{2}$ scales linearly with $N$, which explains the numerical results such as presented in Fig. \ref{fig:PulsePower}. The coupling scales with the number of spins in the sample, but instead of inserting one big magnet, the spins are allowed to be distributed when strongly coupled by the magnon-photon interaction. This process is especially efficient when the magnets are on the chiral line, because all pumped energy is focused into just one propagating cavity mode.

\section{Discussion}

Here we address an ensemble of magnets on the chiral planes of microwave cavities and find that they act as ``photonic wheels'' \cite{banzer_photonic_2013,aiello_transverse_2015} that propel a unidirectional photon spin, momentum, and energy current, but in the microwave regime and on a macroscopic scale. Loading the cavity with \(N\) magnets enhances the effective coupling constant by \(\sqrt N\) while the microwave power at resonance scales like \(N\). The distributed spins are coherently coupled and focus the distributed power input into a single mode and direction. A chiral photon energy current can also be generated by nanomechanical elements arranged on a ring that are coupled and driven optically, while the current direction is controlled by  side-band Raman scattering  \cite{denis_permanent_2020}.


Our torus cavity is representative for other shapes that guide photons into closed orbits such as disks or  (normal of superconducting) co-planar wave guides  \cite{yang_control_2019, li_strong_2019,hou_strong_2019}. An open structure reduces the quality factor, but allows for other detection strategies of chiral microwave currents. For example, the toroidal radiation from the dynamical anapole (see Appendix  \ref{Anapole} can be detected in an open cavity, such as a ring-shaped coplanar waveguide \cite{shao_spin-orbit_2018}.

We may interpret the ring-like cavity with magnets on chiral lines also as a miniature storage ring for the generation of intense polarized microwaves by pumping with distributed weak power sources. Scattering and absorption experiments can be carried out by inserting ``samples'' into designated locations or by coupling the microwaves out of the ring by semitransparent mirrors.

Similar to the aromatic molecules in
conventional chemistry, the collective dynamics of magnets with chiral coupling can be treated as a chiral magnonic molecule, which demonstrates circulating
photonic spin currents, facilitating the design of non-reciprocal devices for
present microwave as well as future quantum information technologies.

\begin{acknowledgments}
	This work was supported by the JSPS Kakenhi (Grant No. 19H006450 and No. 20K14369) and the Nederlandse Organisatie voor Wetenschappelijk Onderzoek (NWO). T. Y. acknowledges the hospitality and support from the IMR.
\end{acknowledgments}

\begin{appendix}

\setcounter{figure}{0}
\setcounter{equation}{0}
\renewcommand\theequation{\Alph{section}\arabic{equation}}
\renewcommand\thefigure{\Alph{section}\arabic{figure}}

\section{Electrodynamics of a torus cavity}\label{Torus}

\begin{figure}[tbh]
\includegraphics[width=8.5cm]{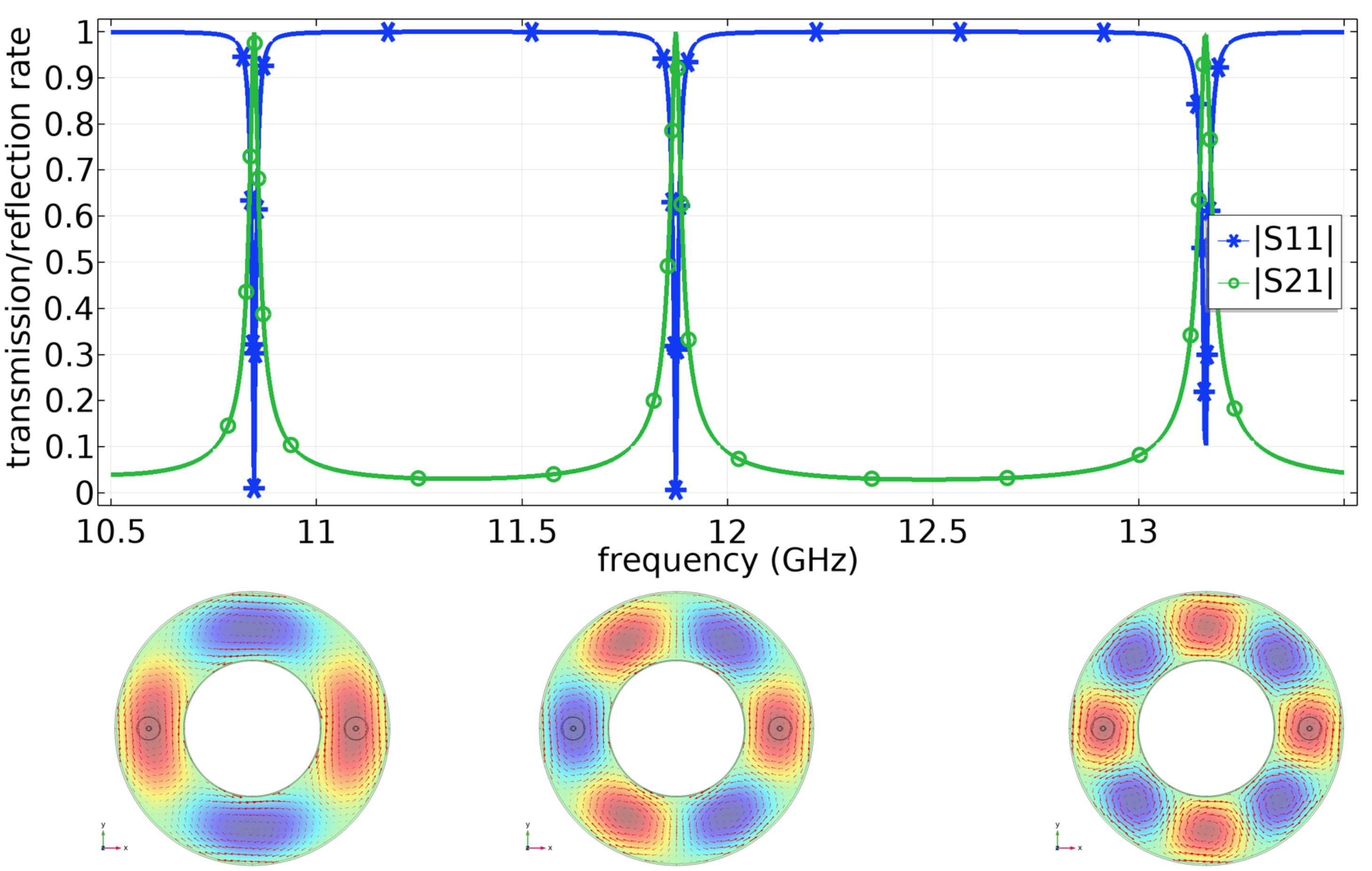}\caption{Numerical reflection (\(S_{11},\) blue) and
transmission ($S_{11}$,green) spectra of the empty cavity. The lowers panels show the electric field amplitudes of the TE modes with $m=\{2,3,4\}$.
\label{fig:S1}}%
\end{figure}

Here we derive the eigenmodes of the torus-shaped cavity with rectangular cross
section (width $w=R_{2}-R_{1}$ and height $h$, where $R_{1/2}$ denotes the
inner/outer radius). The free-space Maxwell equations read
\cite{jackson_classical_1998}
\begin{equation} \nabla\times\mathbf{E}+\mu_0\frac{\partial\mathbf{H}}{\partial t}=0,~~~~~~\nabla\times\mathbf{H}-\varepsilon_0\frac{\partial\mathbf{E}}{\partial t}=0, \label{Maxwell}%
\end{equation}
with (conducting) boundary conditions at metallic cavity walls
\begin{equation}
\mathbf{n}\times\mathbf{E}|_{S}=0,~~~~~~\mathbf{n}\cdot\mathbf{H}|_{S}=0.
\end{equation}
The solutions come in two orthogonal sets, viz. TE-modes with $H_{z}=0$ and
$\partial_{\mathbf{n}}E_{z}|_{S}=0$ and TM-modes with $E_{z}=0$ and
$\partial_{\mathbf{n}}H_{z}|_{S}=0$. Here we focus on the TE modes with
$\mathbf{E}=E_z\hat{\mathbf{z}}$. In cylindrical
coordinates $\{\rho,\phi,z\}$
\begin{equation}
E_{z}=\psi(\rho,\phi)\cos\left(  \frac{p\pi z}{h}\right)  ,~~~~~p=0,1,2,\cdots
.
\end{equation}
with amplitudes \(\psi\)  that obey the wave equation
\begin{equation}
(\nabla_{t}^{2}+\gamma_{p}^{2})\psi_{p}=0,
\end{equation}
where $p$ is the mode number in the \(z\) direction and $\gamma_{p}=\sqrt{\omega^{2}/c^{2}-(p\pi/h)^{2}}$, or
\begin{equation}
\frac{1}{\rho}\frac{\partial}{\partial \rho}\left(  \rho\frac{\partial\psi}{\partial
\rho}\right)  +\frac{1}{\rho^{2}}\frac{\partial^{2}\psi}{\partial\phi^{2}}%
+\gamma_{p}^{2}\psi=0.
\end{equation}
With $\psi(\rho,\phi)=R_{m}(\rho)e^{im\phi}$ and $m\in\mathbb{Z}_{0}$, we
generate the Bessel functions of order $m$ that solve
\cite{li_radial_2012,stupakov_shielding_2003}
\begin{equation}
\frac{1}{\rho}\frac{\partial}{\partial \rho}\left(  \rho\frac{\partial R_{m}%
(\rho)}{\partial \rho}\right)  +\left(  \gamma_{p,m}^{2}-\frac{m^{2}}{\rho^{2}}\right)
R_{m}(\rho)=0,
\end{equation}
The general solution is
\begin{equation}
R_{m}(\rho)=J_{m}(\gamma_{p,m}\rho)+CN_{m}(\gamma_{p,m}\rho),
\end{equation}
with $J_{m}$ and $N_{m}$ being the Bessel functions of the first and second
kind, respectively and the boundary conditions determine the integration constant $C$. At
the inner and outer boundary, the condition with $E_{z}=0$ yields
\begin{align}
J_{m}(\gamma_{p,m}R_{2})+CN_{m}(\gamma_{p,m}R_{2})  &  =0,\\
J_{m}(\gamma_{p,m}R_{1})+CN_{m}(\gamma_{p,m}R_{1})  &  =0.
\end{align}
Thus, $C=-J_{m}(\gamma_{p,m}R_{1})/N_{m}(\gamma_{p,m}R_{1})$ and $\gamma
_{p,m}$ is determined by
\begin{equation}
J_{m}(\gamma_{p,m}R_{2})N_{m}(\gamma_{p,m}R_{1})=J_{m}(\gamma_{p,m}R_{1}%
)N_{m}(\gamma_{p,m}R_{2}). \label{characteristic}%
\end{equation}
The electric field is
\begin{equation}
\begin{aligned} &E_{z}(\rho,\phi,z)=\\ &\left( J_{m}(\gamma_{p,m}\rho)-\frac{J_{m}(\gamma_{p,m} R_{1})}{N_{m}(\gamma_{p,m} R_{1})}N_{m}(\gamma_{p,m}\rho)\right) e^{im\phi}\cos\left( \frac{p\pi z}{h}\right). \end{aligned}
\end{equation}
subject to a normalization factor governed by \Eq{normz}. From the Maxwell equations \ref{Maxwell}
we obtain the magnetic field components of the TE mode
\begin{equation}
\begin{aligned} H_{x}(\rho,\phi,z) & =-i\frac{\omega}{\mu_{0}\gamma_{p,m}^{2}c^{2}}\partial _{y}E_{z}\\ & =-i\frac{\omega}{\mu_{0}\gamma_{p,m}^{2}c^{2}}\left( \sin\phi \frac{\partial}{\partial \rho}+\frac{1}{\rho}\cos\phi\frac{\partial}{\partial\phi}\right) E_{z},\\ H_{y}(\rho,\phi,z) & =i\frac{\omega}{\mu_{0}\gamma_{p,m}^{2}c^{2}}\partial _{x}E_{z}\\ &=i\frac{\omega}{\mu_{0}\gamma_{p,m}^{2}c^{2}}\left( \cos\phi\frac {\partial}{\partial \rho}-\frac{1}{\rho}\sin\phi\frac{\partial}{\partial\phi }\right) E_{z}. \end{aligned}
\end{equation}
For small \(h\) modes with \(p \ne 0\)  are pushed to high frequencies and may be disregarded. In cylindrical coordinates, \(H_{z}=0\) and  $\mathbf{H}%
(\rho,\phi)=H_{\rho}\hat{\mathbf{e}}_{\rho}+H_{\phi}\hat{\mathbf{e}}_{\phi}$ with
\begin{equation}
\begin{aligned} H_{\rho}(\rho,\phi,z) & =\frac{1}{\mu_0\gamma_mc}\frac{m}{\rho}E_z\\ H_{\phi}(\rho,\phi,z) & =-i\frac{1}{\mu_0\gamma_mc}\frac{\partial E_z}{\partial \rho}\\ \end{aligned}
\end{equation}
With $R_{1}=15$~mm and
$R_{2}=30$~mm,  Eq.~(\ref{characteristic}) leads to the eigenfrequencies $\omega_m=c\gamma_{p=0,m}=\{10.84,11.87,13.16\}$ GHz
for $m=\{2,3,4\}$, which agrees with the numerical results as shown in \Figure{fig:S1}.

The field distribution of small torus cavities differs from  straight open waveguides. We observe in Figure \ref{fig1} that the magnetic field is not maximal in the center and at the chiral lines but shifted to the edges. Increasing the torus circumference for constant cross section shifts the maximum field into the center, but also decreases the local amplitude of a given cavity mode. The straight wave guide limit is reached when the discrete \(m\)-modes merge into a one-dimensional continuum.

\section{Numerical method}\label{Numerics}

We conduct finite-element simulations based on COMSOL Multiphysics
\cite{comsol}. The spin dynamics is governed by the  Landau-Lifshitz-Gilbert
equation
\begin{equation}
\frac{\partial\mathbf{m}}{\partial t}=-\gamma\mathbf{m}\times\mathbf{H}
_{\text{eff}}+\alpha\mathbf{m}\times\frac{\partial\mathbf{m}}{\partial t}
\label{eqn:LLG}%
\end{equation}
with (modulus of) the gyromagnetic ratio $\gamma$, the Gilbert damping
constant $\alpha$ and the  effective field $\mathbf{H}_{\text{eff}}$
containing the exchange interaction,  static external field $H_{0}$, and
dynamic magnetic field $\mathbf{H}$ (including the dipolar field
$\mathbf{H}_{d}$ inside the sphere). The magnetic field inside and outside the
sphere is the solution of Maxwell's equation. The vector potential
$\mathbf{A}(\mathbf{r},t)$ that governs the  dynamics of the electromagnetic waves%

\begin{equation}
\mu_{0}\varepsilon_{0}\frac{\partial}{\partial t}\left(  \varepsilon
_{r}(\mathbf{r}) \frac{\partial\mathbf{A}}{\partial t}\right)  +\nabla
\times\left(  \mu_{r}^{-1}(\mathbf{r})\nabla\times\mathbf{A}\right)  =0
\label{eqn:Maxwell}%
\end{equation}
where the relative permittivity $\varepsilon_{r}=1(15)$ \cite{sadhana_synthesis_2009} outside (inside) the
sphere and the relative permeability $\mu_{r}(\mathbf{r})=1$ for both inside
and outside  the sphere. The standard magnetic boundary condition at the
surface of the sphere reads
\begin{equation}
\begin{aligned} \mathbf{n}\times\left(\frac{\partial\mathbf{A}_1}{\partial t}-\frac{\partial\mathbf{A}_2}{\partial t}\right)&=0\\ \mathbf{n}\cdot\left(\varepsilon_1\frac{\partial\mathbf{A}_1}{\partial t}-\varepsilon_2\frac{\partial\mathbf{A}_2}{\partial t}\right)&=0\\ \mathbf{n}\times(\frac{1}{\mu_0}\nabla\times\mathbf{A}_1-\frac{1}{\mu_2}\nabla\times\mathbf{A}_2-\mathbf{M})&=0\\ \mathbf{n}\cdot\left(\nabla\times\mathbf{A}_1-\nabla\times\mathbf{A}_2\right)&=0, \end{aligned}
\end{equation}
where the label $1,2$ denotes the region inside and outside the sphere. The last line corresponds to $\mathbf{n}\cdot(\mathbf{B}_1-\mathbf{B}_2)=0$, and $\mathbf{B}=\mu_0(\mathbf{H}+\mathbf{M})$. The solutions of these coupled equation entail the dynamics of the hybrid system.

In the first approach, we solve the linearized coupled equations in
the  frequency domain. The magnetization which in the ground state is along the $\hat{\mathbf{z}}$ direction  can be written as $\mathbf{M}= M_{s}\hat{\mathbf{z}}+\mathbf{m}$,
where the small amplitude $\mathbf{m}$ satisfies:%

\begin{equation}
i\omega\mathbf{m}=\hat{\mathbf{z}}\times(\omega_{M}\mathbf{H}-\omega
_{\rm{K}}\mathbf{m}+i\alpha\omega\mathbf{m})
\end{equation}
where $\omega_{M}=\gamma M_{s}$ with the saturated magnetization
$M_{s}=\SI{0.176} \,\mathrm{{T}}$ \cite{wang_magnon_2016}, gyromagnetic ratio
$\gamma/(2\pi)=\SI{28} \, \mathrm{{GHz/T}}$, the magnon dissipation rate
$\kappa=2\alpha\omega_{\rm{K}}$ with Gilbert damping $\alpha=7.8\times10^{-5}$.  In
the cavity, the Kittel mode frequency $\omega_{\rm{K}}=\gamma H_{0}$ is slightly
shifted by about  \SI{56}{MHz} for the considered configuration. Since
\Eq{eqn:LLG} can be  represented in the form $\mathbf{m}=\overline{\zeta}\mathbf{h}$, and
the equivalent permeability of the  magnetic sphere is
\cite{zare_rameshti_indirect_2018,cao_exchange_2015}
\begin{equation}
\overline{\mu}_{\text{M}}=\overline{\text{I}}+\overline{\zeta}=\left(
\begin{array}
[c]{ccc}%
1+u & -iv & 0\\
iv & 1+u & 0\\
0 & 0 & 1
\end{array}
\right)
\end{equation}
where
\begin{equation}
u=\frac{\left(  \omega_{K}-i\alpha\omega\right)  \omega_{M}}{\left(
\omega_{K}-i\alpha\omega\right)  ^{2}-\omega^{2}},\quad v=\frac{\omega
\omega_{M}}{\left(  \omega_{K}-i\alpha\omega\right)  ^{2}-\omega^{2}}%
\end{equation}
The electromagnetic wave in the cavity (including the magnetic sphere) is
governed by the Maxwell's equation
\begin{equation}
\nabla\times[\mu_{r}(\mathbf{r})^{-1}\nabla\times\mathbf{E}]-k^{2}%
\varepsilon_{r}(\mathbf{r})\mathbf{E}=0
\end{equation}
where $k=\omega/c$ is the wave vector of light in vacuum. The relative permeability $\mu_{r}(\mathbf{r})=1$
outside the magnetic sphere,  and $\mu_{r}(\mathbf{r})=\overline{\mu}_{\text{M}}$ inside
the magnetic sphere. The boundary condition at the surface of the sphere
reads
\begin{equation}
\begin{aligned} \mathbf{n}\times\left(\mathbf{E}_1-\mathbf{E}_2\right)&=0\\ \mathbf{n}\cdot\left(\varepsilon_1\mathbf{E}_1-\varepsilon_2\mathbf{E}_2\right)&=0\\ \mathbf{n}\times\left(\mathbf{H}_1-\mathbf{H}_2\right)&=0\\ \mathbf{n}\cdot\left(\mu_1\mathbf{H}_1-\mu_2\mathbf{H}_2\right)&=0 \end{aligned}
\end{equation}
where labels $1,2$ denotes the region inside and outside the sphere. Since the
magnetic dynamics is treated by an effective permeability, this code does not
capture the static stray field from the magnetic sphere, which is not an issue
as long as different spheres are sufficiently separated from each other. By
connecting two  ports with $(\varepsilon_{r},\mu_{r})=(2,1)$ to the system and
integrate the electric field on the surface of the ports, we can calculate the
transmission (reflection) spectrum  $S_{21}$ ($S_{11}$) \cite{comsol}.

In the time-dependent simulations we take into account the full
non-linearity of the LLG, but use the macrospin approximation for the magnetization field. A magnetic field pulse is locally applied on the spheres to mimic the excitation by proximity coils as used in experiments \cite{zhang_magnon_2015,lambert_cavity-mediated_2016}.

\section{Quantization of the electromagnetic field}\label{QPhoton}

The photon Hamiltonian is
\begin{equation}
\hat{H}_{p}=\int d\mathbf{r}\left[  \frac{\varepsilon_{0}}{2}\mathbf{E}
(\mathbf{r})\cdot\mathbf{E}(\mathbf{r})+\frac{\mu_{0}}{2}\mathbf{H}
(\mathbf{r})\cdot\mathbf{H}(\mathbf{r})\right]  . \label{photon_Hamiltonian}%
\end{equation}
For our TE mode, the electric and magnetic fields are quantized by the photon
operator $\hat{\alpha}_{p,m}$,
\begin{align}
\mathbf{H}(\mathbf{r})  &  =\sum_{p}\sum_{m}\left[  {\mathcal H}_{m}^{p}
(\rho,z)e^{im\phi}\hat{\alpha}_{p,m}+{\mathcal H}_{m}^{p\ast}(\rho
,z)e^{-im\phi}\hat{\alpha}_{p,m}^{\dagger}\right]  ,\nonumber\\
\mathbf{E}(\mathbf{r})  &  =\sum_{p}\sum_{m}\left[  {\mathcal E}_{m}^{p}
(\rho,z)e^{im\phi}\hat{\alpha}_{p,m}+{\mathcal E}_{m}^{p\ast}(\rho
,z)e^{-im\phi}\hat{\alpha}_{p,m}^{\dagger}\right]  ,
\label{quantization_photon}%
\end{align}
which can quantize Eq.~(\ref{photon_Hamiltonian}) to the harmonic oscillator
$\hat{H}_{p}=\sum_{p,m}\hbar\omega_{p,m}\hat{\alpha}_{p,m}^{\dagger}
\hat{\alpha}_{p,m}$ with the normalization condition
\begin{equation}
\int d\mathbf{r}\left(  \frac{\varepsilon_{0}}{2}|{\mathcal E}_{m}^{p}
(\rho,z)|^{2}+\frac{\mu_{0}}{2}|{\mathcal H}_{m}^{p}(\rho,z)|^{2}\right)
=\frac{\hbar\omega_{p,m}}{2},
\end{equation}
leading to
\begin{equation}
\pi\int_{R_{1}}^{R_{2}}\rho d\rho\int_{0}^{b}dz\left(  \varepsilon_{0}
|{\mathcal E}_{m}^{p}(\rho,z)|^{2}+\mu_{0}|{\mathcal H}_{m}^{p}(\rho
,z)|^{2}\right)  =\frac{\hbar\omega_{p,m}}{2}.
\label{normz}
\end{equation}
We focus on TE mode with $p=0$ in this work.

\section{Quantum Hamiltonian for multiple magnets}\label{MM}

By loading the torus-like cavity with $N$ small magnetic spheres centered at
positions $\{(\rho_{l},\phi_{l})\}$ ($l=1,2,\cdots N$), the photon can couple
to the magnetization $M_{l}(r)$ through the Zeeman interaction
\begin{equation}
\hat{H}_{\mathrm{int}}=-{\mu_{0}}\sum_{l=1}^{N}\int d\mathbf{r}\left[
\mathbf{H}_{0}^{\left(  l\right)  }(\mathbf{r})\cdot\mathbf{M}
_{l}(\mathbf{r})+\mathbf{H}(\mathbf{r})\cdot\mathbf{M}_{l}(\mathbf{r})\right]
,
\end{equation}
where $H_{0}^{\left(  l\right)  }$ represents the applied static
magnetic field that saturates the magnetization of each sphere to $M_{l,s}$.
The magnetization can be expressed by the spin operator as $M_{l}
(\mathbf{r})=-\gamma\hbar S_{l}(\mathbf{r})$, where $-\gamma$ is the gyromagnetic ratio. The
microwaves with amplitude nearly constant over the magnets excite only the
uniform precession of the ordered spins, i.e. the Kittel mode.

With saturated magnetization along the $\hat{\mathbf{z}}$-direction, the spin
operators can be expanded by the Kittel mode as
\begin{equation}
\hat{S}_{l,\beta}^{\mathrm{K}}(\mathbf{r},t)=\sqrt{2M_{l,s}/(\gamma\hbar
)}\left[  M_{l,\beta}^{\mathrm{K}}(\mathbf{r})\hat{m}_{l}(t)+M_{l,\beta
}^{\mathrm{K}\ast}(\mathbf{r})\hat{m}_{l}^{\dagger}(t)\right]  ,
\label{quantization_magnon}%
\end{equation}
in which $\hat{m}_{l}$ denotes the magnon annihilation operator, and
$M_{l,\beta}^{K}$ is the corresponding eigenfunction of the Kittel mode
(which  is constant inside and zero outside the magnet) with $\beta=\{x,y\}$
and  normalization
\begin{equation}
\int d\boldsymbol{\rho}dz\left[  M_{l,x}^{\mathrm{K}}(\mathbf{r}
)M_{l,y}^{\mathrm{K}\ast}(\mathbf{r})-M_{l,x}^{\mathrm{K}\ast}(\mathbf{r}
)M_{l,y}^{\mathrm{K}}(\mathbf{r})\right]  =-i/2.
\end{equation}
Because the magnetization is uniform in the magnetic sphere and the Kittel
mode is circularly polarized with $M_{l,y}=iM_{l,x}$. By Eqs.~(\ref{quantization_magnon}) and (\ref{quantization_photon}), the
interaction Hamiltonian in the rotating wave approximation reads
\begin{equation}
\hat{H}_{\mathrm{int}}=\hbar\sum_{m}\sum_{l=1}^{N}|g_{l,m}|e^{im_\ast\phi_l}\hat{\alpha}_{m}%
\hat{m}_{l}^{\dagger}+\mathrm{H.c.},
\end{equation}
where $m_\ast=m+1$ due to the topology of the torus-shaped cavity and coupling strength \cite{yu_magnon_2020}
\begin{align}
g_{l,m} ={\mu_{0}}\sqrt{\frac{\gamma M_{l,s}V_{l,s}}{2\hbar}}\left[\mathcal{H}
_{\rho}^{m}({\rho}_{l})+i\mathcal{H}_{\phi}^{m}({\rho}_{l})\right].\label{coupling}%
\end{align}
 where $V_{l,s}$ is the volume of each magnetic sphere, $\mathcal{H}_{\rho}^{m}({\rho}_{l})$ and $\mathcal{H}_{\phi}^{m}({\rho}_{l})$ are the normalized magnetic field calculated according to \Eq{normz}. Figure \ref{fig1}(b) in the main text shows the coupling strength according to \Eq{coupling} for $m=\pm2$. In general, $g_{m}\neq g_{-m}$, implying at least partially-chiral coupling,
depending on the position of the sphere, i.e., the coupling strength between
magnon and photon can depend on the photon propagation direction. We disregard the normal scattering $\hbar g_{n}\sum_{m}\hat{\alpha
}_{m}^{\dagger}\hat{\alpha}_{-m}$ since the simulations show that they cause only weak effects in the present set-up.

\section{Anapole}\label{Anapole}

The anapole (toroidal dipole moment) is a localized electromagnetic excitation, distinct from the magnetic and electric dipoles. The anapole is defined in terms of the far field of a torus with electric currents flowing on its surface as the leading term in a toroidal multipole expansion. \cite{papasimakis_electromagnetic_2016,talebi_theory_2018}. Alternatively, static and dynamic magnetizations contribute to the anapole  \cite{ederer_towards_2007}
\begin{equation}
\mathbf{T}=\frac{1}{2}\sum_{j}\mathbf{r}_j\times\mathbf{M}_j
\label{eqn:anapole}
\end{equation}
The magnonic molecule is composed of spatially distributed magnets that can contribute a static anapole \(\mathbf{T}_\text{S}\). A \(\mathbf{T}_\text{D}\) is generated by the dynamical magnetization $\mathbf{m}=\mathbf{M}-\mathbf{M}_0$, where $\mathbf{M}_0$ is the equilibrium magnetization.

The static anapole of a ring of magnets vanishes when magnetizations are parallel and is maximized for an in-plane closed-flux configuration, \eg $\mathbf{M}_0^j= \pm M_s(-\cos\phi_j,\sin\phi_j,0)$ at positions $\mathbf{r}_j=\rho(\sin\phi_j,\cos\phi_j,0)$:
\begin{equation}
\mathbf{T}_{\rm{S}}
=\pm\frac{1}{2}\rho M_s\sum_j(\sin^2\phi_j+\cos^2\phi_j)\hat{\mathbf{z}}
=\pm\frac{N}{2}\rho M_s\hat{\mathbf{z}}
\label{eqn:staticanapole}
\end{equation}
where $\pm$ indicates the direction of the moments.

When the spins interact, the static anapole corresponds to a strained magnetic configuration, which is equivalent to a ground state spin current \(\mathbf{J}_s=-\delta E[\mathbf{M}]/\delta \mathbf{M}\). The dipolar energy density of the $i$-th magnet is $E_i=-\mu_0\mathbf{M}_i\cdot\sum_j\mathbf{H}_d$, where $\mathbf{H}_d$ is the dipolar field contributed by sphere $j$ at the position of sphere $i$. The spin current vanishes when the system minimizes or maximizes the energy, i.e. when all magnetizations are collinear. In that limit $\mathbf{T}_{\rm{S}}$ vanishes. The concept of  \( \mathbf{T}_{\rm{S}}\) is derived from the dipolar energy in vacuum, so it cannot be related to cavity-induced interactions between the magnets.

The transverse collective motion of the magnonic molecule also contributes, however. For $N$ evenly distributed spheres on the radial position $\rho$, precessing like $\mathbf{m}_j(t)=m_0\exp[i\omega t+i\varphi_j]$ and equal average amplitude $m_0$  the dynamical anapole reads
\begin{align}
\mathbf{T}_\text{D}&=\frac{1}{2}\sum_{j}\mathbf{r}_j\times\mathbf{m}_j(t)\nonumber\\
&=\frac{\rho m_0 }{2}\sum_{j}\left(\sin\phi_j\cos(\omega t+\varphi_j)-\cos\phi_j\sin(\omega t+\varphi_j)\right)\hat{\mathbf{z}}\nonumber\\
&=\frac{\rho m_0}{2}\sum_{j}\sin\left(\phi_j-\varphi_j-\omega t\right)\hat{\mathbf{z}}.
\end{align}
Depending on the polar position $\phi_j$ and phase $\varphi_j$, the dynamical anapole oscillates in time or vanishes. For the magnonic homo-dimer ($N=2$ on $\rho_+$, and $m=2$)  $\mathbf{T}_\text{D}=\rho_+m_0e^{i(\omega_{\mathrm K}+2\Gamma_{+})t}\hat{\mathbf{z}}$  in the bright and vanishes for the dark state. When $N=3$, however, the roles of bright and dark states are reversed. In a closed cavity this difference is inconsequential, but in an open cavity the anapole is associated with toroidal microwave generation \cite{heras_electric_1998,talebi_theory_2018} which can be easily controlled here. We also see that in open cavities the dark states are dark only in the dipolar field, but radiate energy into toroidal channels. A uni-direction of the microwave flux does not affect the anapole directly, however.

\end{appendix}


\begin{thebibliography}{53}%
\makeatletter
\providecommand \@ifxundefined [1]{%
 \@ifx{#1\undefined}
}%
\providecommand \@ifnum [1]{%
 \ifnum #1\expandafter \@firstoftwo
 \else \expandafter \@secondoftwo
 \fi
}%
\providecommand \@ifx [1]{%
 \ifx #1\expandafter \@firstoftwo
 \else \expandafter \@secondoftwo
 \fi
}%
\providecommand \natexlab [1]{#1}%
\providecommand \enquote  [1]{``#1''}%
\providecommand \bibnamefont  [1]{#1}%
\providecommand \bibfnamefont [1]{#1}%
\providecommand \citenamefont [1]{#1}%
\providecommand \href@noop [0]{\@secondoftwo}%
\providecommand \href [0]{\begingroup \@sanitize@url \@href}%
\providecommand \@href[1]{\@@startlink{#1}\@@href}%
\providecommand \@@href[1]{\endgroup#1\@@endlink}%
\providecommand \@sanitize@url [0]{\catcode `\\12\catcode `\$12\catcode
  `\&12\catcode `\#12\catcode `\^12\catcode `\_12\catcode `\%12\relax}%
\providecommand \@@startlink[1]{}%
\providecommand \@@endlink[0]{}%
\providecommand \url  [0]{\begingroup\@sanitize@url \@url }%
\providecommand \@url [1]{\endgroup\@href {#1}{\urlprefix }}%
\providecommand \urlprefix  [0]{URL }%
\providecommand \Eprint [0]{\href }%
\providecommand \doibase [0]{http://dx.doi.org/}%
\providecommand \selectlanguage [0]{\@gobble}%
\providecommand \bibinfo  [0]{\@secondoftwo}%
\providecommand \bibfield  [0]{\@secondoftwo}%
\providecommand \translation [1]{[#1]}%
\providecommand \BibitemOpen [0]{}%
\providecommand \bibitemStop [0]{}%
\providecommand \bibitemNoStop [0]{.\EOS\space}%
\providecommand \EOS [0]{\spacefactor3000\relax}%
\providecommand \BibitemShut  [1]{\csname bibitem#1\endcsname}%
\let\auto@bib@innerbib\@empty
\bibitem [{\citenamefont {Zhang}\ \emph {et~al.}(2014)\citenamefont {Zhang},
  \citenamefont {Zou}, \citenamefont {Jiang},\ and\ \citenamefont
  {Tang}}]{zhang_strongly_2014}%
  \BibitemOpen
  \bibfield  {author} {\bibinfo {author} {\bibfnamefont {Xufeng}\ \bibnamefont
  {Zhang}}, \bibinfo {author} {\bibfnamefont {Chang-Ling}\ \bibnamefont {Zou}},
  \bibinfo {author} {\bibfnamefont {Liang}\ \bibnamefont {Jiang}}, \ and\
  \bibinfo {author} {\bibfnamefont {Hong~X.}\ \bibnamefont {Tang}},\ }\bibfield
   {title} {\enquote {\bibinfo {title} {Strongly {Coupled} {Magnons} and
  {Cavity} {Microwave} {Photons}},}\ }\href {\doibase
  10.1103/PhysRevLett.113.156401} {\bibfield  {journal} {\bibinfo  {journal}
  {Physical Review Letters}\ }\textbf {\bibinfo {volume} {113}},\ \bibinfo
  {pages} {156401} (\bibinfo {year} {2014})}\BibitemShut {NoStop}%
\bibitem [{\citenamefont {Huebl}\ \emph {et~al.}(2013)\citenamefont {Huebl},
  \citenamefont {Zollitsch}, \citenamefont {Lotze}, \citenamefont {Hocke},
  \citenamefont {Greifenstein}, \citenamefont {Marx}, \citenamefont {Gross},\
  and\ \citenamefont {Goennenwein}}]{huebl_high_2013}%
  \BibitemOpen
  \bibfield  {author} {\bibinfo {author} {\bibfnamefont {Hans}\ \bibnamefont
  {Huebl}}, \bibinfo {author} {\bibfnamefont {Christoph~W.}\ \bibnamefont
  {Zollitsch}}, \bibinfo {author} {\bibfnamefont {Johannes}\ \bibnamefont
  {Lotze}}, \bibinfo {author} {\bibfnamefont {Fredrik}\ \bibnamefont {Hocke}},
  \bibinfo {author} {\bibfnamefont {Moritz}\ \bibnamefont {Greifenstein}},
  \bibinfo {author} {\bibfnamefont {Achim}\ \bibnamefont {Marx}}, \bibinfo
  {author} {\bibfnamefont {Rudolf}\ \bibnamefont {Gross}}, \ and\ \bibinfo
  {author} {\bibfnamefont {Sebastian T.~B.}\ \bibnamefont {Goennenwein}},\
  }\bibfield  {title} {\enquote {\bibinfo {title} {High {Cooperativity} in
  {Coupled} {Microwave} {Resonator} {Ferrimagnetic} {Insulator} {Hybrids}},}\
  }\href {\doibase 10.1103/PhysRevLett.111.127003} {\bibfield  {journal}
  {\bibinfo  {journal} {Physical Review Letters}\ }\textbf {\bibinfo {volume}
  {111}},\ \bibinfo {pages} {127003} (\bibinfo {year} {2013})}\BibitemShut
  {NoStop}%
\bibitem [{\citenamefont {Tabuchi}\ \emph {et~al.}(2014)\citenamefont
  {Tabuchi}, \citenamefont {Ishino}, \citenamefont {Ishikawa}, \citenamefont
  {Yamazaki}, \citenamefont {Usami},\ and\ \citenamefont
  {Nakamura}}]{tabuchi_hybridizing_2014}%
  \BibitemOpen
  \bibfield  {author} {\bibinfo {author} {\bibfnamefont {Yutaka}\ \bibnamefont
  {Tabuchi}}, \bibinfo {author} {\bibfnamefont {Seiichiro}\ \bibnamefont
  {Ishino}}, \bibinfo {author} {\bibfnamefont {Toyofumi}\ \bibnamefont
  {Ishikawa}}, \bibinfo {author} {\bibfnamefont {Rekishu}\ \bibnamefont
  {Yamazaki}}, \bibinfo {author} {\bibfnamefont {Koji}\ \bibnamefont {Usami}},
  \ and\ \bibinfo {author} {\bibfnamefont {Yasunobu}\ \bibnamefont
  {Nakamura}},\ }\bibfield  {title} {\enquote {\bibinfo {title} {Hybridizing
  {Ferromagnetic} {Magnons} and {Microwave} {Photons} in the {Quantum}
  {Limit}},}\ }\href {\doibase 10.1103/PhysRevLett.113.083603} {\bibfield
  {journal} {\bibinfo  {journal} {Physical Review Letters}\ }\textbf {\bibinfo
  {volume} {113}},\ \bibinfo {pages} {083603} (\bibinfo {year}
  {2014})}\BibitemShut {NoStop}%
\bibitem [{\citenamefont {Chumak}\ \emph {et~al.}(2015)\citenamefont {Chumak},
  \citenamefont {Vasyuchka}, \citenamefont {Serga},\ and\ \citenamefont
  {Hillebrands}}]{chumak_magnon_2015}%
  \BibitemOpen
  \bibfield  {author} {\bibinfo {author} {\bibfnamefont {A.~V.}\ \bibnamefont
  {Chumak}}, \bibinfo {author} {\bibfnamefont {V.~I.}\ \bibnamefont
  {Vasyuchka}}, \bibinfo {author} {\bibfnamefont {A.~A.}\ \bibnamefont
  {Serga}}, \ and\ \bibinfo {author} {\bibfnamefont {B.}~\bibnamefont
  {Hillebrands}},\ }\bibfield  {title} {{\selectlanguage {english}\enquote
  {\bibinfo {title} {Magnon spintronics},}\ }}\href {\doibase
  10.1038/nphys3347} {\bibfield  {journal} {\bibinfo  {journal} {Nature
  Physics}\ }\textbf {\bibinfo {volume} {11}},\ \bibinfo {pages} {453}
  (\bibinfo {year} {2015})}\BibitemShut {NoStop}%
\bibitem [{\citenamefont {Bai}\ \emph {et~al.}(2015)\citenamefont {Bai},
  \citenamefont {Harder}, \citenamefont {Chen}, \citenamefont {Fan},
  \citenamefont {Xiao},\ and\ \citenamefont {Hu}}]{bai_spin_2015}%
  \BibitemOpen
  \bibfield  {author} {\bibinfo {author} {\bibfnamefont {Lihui}\ \bibnamefont
  {Bai}}, \bibinfo {author} {\bibfnamefont {M.}~\bibnamefont {Harder}},
  \bibinfo {author} {\bibfnamefont {Y.~P.}\ \bibnamefont {Chen}}, \bibinfo
  {author} {\bibfnamefont {X.}~\bibnamefont {Fan}}, \bibinfo {author}
  {\bibfnamefont {J.~Q.}\ \bibnamefont {Xiao}}, \ and\ \bibinfo {author}
  {\bibfnamefont {C.-M.}\ \bibnamefont {Hu}},\ }\bibfield  {title} {\enquote
  {\bibinfo {title} {Spin {Pumping} in {Electrodynamically} {Coupled}
  {Magnon}-{Photon} {Systems}},}\ }\href {\doibase
  10.1103/PhysRevLett.114.227201} {\bibfield  {journal} {\bibinfo  {journal}
  {Physical Review Letters}\ }\textbf {\bibinfo {volume} {114}},\ \bibinfo
  {pages} {227201} (\bibinfo {year} {2015})}\BibitemShut {NoStop}%
\bibitem [{\citenamefont {Maier-Flaig}\ \emph {et~al.}(2016)\citenamefont
  {Maier-Flaig}, \citenamefont {Harder}, \citenamefont {Gross}, \citenamefont
  {Huebl},\ and\ \citenamefont {Goennenwein}}]{maier-flaig_spin_2016}%
  \BibitemOpen
  \bibfield  {author} {\bibinfo {author} {\bibfnamefont {H.}~\bibnamefont
  {Maier-Flaig}}, \bibinfo {author} {\bibfnamefont {M.}~\bibnamefont {Harder}},
  \bibinfo {author} {\bibfnamefont {R.}~\bibnamefont {Gross}}, \bibinfo
  {author} {\bibfnamefont {H.}~\bibnamefont {Huebl}}, \ and\ \bibinfo {author}
  {\bibfnamefont {S.~T.~B.}\ \bibnamefont {Goennenwein}},\ }\bibfield  {title}
  {\enquote {\bibinfo {title} {Spin pumping in strongly coupled magnon-photon
  systems},}\ }\href {\doibase 10.1103/PhysRevB.94.054433} {\bibfield
  {journal} {\bibinfo  {journal} {Physical Review B}\ }\textbf {\bibinfo
  {volume} {94}},\ \bibinfo {pages} {054433} (\bibinfo {year}
  {2016})}\BibitemShut {NoStop}%
\bibitem [{\citenamefont {Bai}\ \emph {et~al.}(2017)\citenamefont {Bai},
  \citenamefont {Harder}, \citenamefont {Hyde}, \citenamefont {Zhang},
  \citenamefont {Hu}, \citenamefont {Chen},\ and\ \citenamefont
  {Xiao}}]{bai_cavity_2017}%
  \BibitemOpen
  \bibfield  {author} {\bibinfo {author} {\bibfnamefont {Lihui}\ \bibnamefont
  {Bai}}, \bibinfo {author} {\bibfnamefont {Michael}\ \bibnamefont {Harder}},
  \bibinfo {author} {\bibfnamefont {Paul}\ \bibnamefont {Hyde}}, \bibinfo
  {author} {\bibfnamefont {Zhaohui}\ \bibnamefont {Zhang}}, \bibinfo {author}
  {\bibfnamefont {Can-Ming}\ \bibnamefont {Hu}}, \bibinfo {author}
  {\bibfnamefont {Y.~P.}\ \bibnamefont {Chen}}, \ and\ \bibinfo {author}
  {\bibfnamefont {John~Q.}\ \bibnamefont {Xiao}},\ }\bibfield  {title}
  {\enquote {\bibinfo {title} {Cavity {Mediated} {Manipulation} of {Distant}
  {Spin} {Currents} {Using} a {Cavity}-{Magnon}-{Polariton}},}\ }\href
  {\doibase 10.1103/PhysRevLett.118.217201} {\bibfield  {journal} {\bibinfo
  {journal} {Physical Review Letters}\ }\textbf {\bibinfo {volume} {118}},\
  \bibinfo {pages} {217201} (\bibinfo {year} {2017})}\BibitemShut {NoStop}%
\bibitem [{\citenamefont {Wang}\ \emph {et~al.}(2019)\citenamefont {Wang},
  \citenamefont {Rao}, \citenamefont {Yang}, \citenamefont {Xu}, \citenamefont
  {Gui}, \citenamefont {Yao}, \citenamefont {You},\ and\ \citenamefont
  {Hu}}]{wang_nonreciprocity_2019}%
  \BibitemOpen
  \bibfield  {author} {\bibinfo {author} {\bibfnamefont {Yi-Pu}\ \bibnamefont
  {Wang}}, \bibinfo {author} {\bibfnamefont {J.~W.}\ \bibnamefont {Rao}},
  \bibinfo {author} {\bibfnamefont {Y.}~\bibnamefont {Yang}}, \bibinfo {author}
  {\bibfnamefont {Peng-Chao}\ \bibnamefont {Xu}}, \bibinfo {author}
  {\bibfnamefont {Y.~S.}\ \bibnamefont {Gui}}, \bibinfo {author} {\bibfnamefont
  {B.~M.}\ \bibnamefont {Yao}}, \bibinfo {author} {\bibfnamefont {J.~Q.}\
  \bibnamefont {You}}, \ and\ \bibinfo {author} {\bibfnamefont {C.-M.}\
  \bibnamefont {Hu}},\ }\bibfield  {title} {\enquote {\bibinfo {title}
  {Nonreciprocity and {Unidirectional} {Invisibility} in {Cavity}
  {Magnonics}},}\ }\href {\doibase 10.1103/PhysRevLett.123.127202} {\bibfield
  {journal} {\bibinfo  {journal} {Physical Review Letters}\ }\textbf {\bibinfo
  {volume} {123}},\ \bibinfo {pages} {127202} (\bibinfo {year}
  {2019})}\BibitemShut {NoStop}%
\bibitem [{\citenamefont {Rao}\ \emph {et~al.}(2019)\citenamefont {Rao},
  \citenamefont {Kaur}, \citenamefont {Yao}, \citenamefont {Edwards},
  \citenamefont {Zhao}, \citenamefont {Fan}, \citenamefont {Xue}, \citenamefont
  {Silva}, \citenamefont {Gui},\ and\ \citenamefont {Hu}}]{rao_analogue_2019}%
  \BibitemOpen
  \bibfield  {author} {\bibinfo {author} {\bibfnamefont {J.~W.}\ \bibnamefont
  {Rao}}, \bibinfo {author} {\bibfnamefont {S.}~\bibnamefont {Kaur}}, \bibinfo
  {author} {\bibfnamefont {B.~M.}\ \bibnamefont {Yao}}, \bibinfo {author}
  {\bibfnamefont {E.~R.~J.}\ \bibnamefont {Edwards}}, \bibinfo {author}
  {\bibfnamefont {Y.~T.}\ \bibnamefont {Zhao}}, \bibinfo {author}
  {\bibfnamefont {Xiaolong}\ \bibnamefont {Fan}}, \bibinfo {author}
  {\bibfnamefont {Desheng}\ \bibnamefont {Xue}}, \bibinfo {author}
  {\bibfnamefont {T.~J.}\ \bibnamefont {Silva}}, \bibinfo {author}
  {\bibfnamefont {Y.~S.}\ \bibnamefont {Gui}}, \ and\ \bibinfo {author}
  {\bibfnamefont {C.-M.}\ \bibnamefont {Hu}},\ }\bibfield  {title}
  {{\selectlanguage {english}\enquote {\bibinfo {title} {Analogue of dynamic
  {Hall} effect in cavity magnon polariton system and coherently controlled
  logic device},}\ }}\href {\doibase 10.1038/s41467-019-11021-2} {\bibfield
  {journal} {\bibinfo  {journal} {Nature Communications}\ }\textbf {\bibinfo
  {volume} {10}},\ \bibinfo {pages} {1--7} (\bibinfo {year}
  {2019})}\BibitemShut {NoStop}%
\bibitem [{\citenamefont {Zhang}\ \emph {et~al.}(2015)\citenamefont {Zhang},
  \citenamefont {Zou}, \citenamefont {Zhu}, \citenamefont {Marquardt},
  \citenamefont {Jiang},\ and\ \citenamefont {Tang}}]{zhang_magnon_2015}%
  \BibitemOpen
  \bibfield  {author} {\bibinfo {author} {\bibfnamefont {Xufeng}\ \bibnamefont
  {Zhang}}, \bibinfo {author} {\bibfnamefont {Chang-Ling}\ \bibnamefont {Zou}},
  \bibinfo {author} {\bibfnamefont {Na}~\bibnamefont {Zhu}}, \bibinfo {author}
  {\bibfnamefont {Florian}\ \bibnamefont {Marquardt}}, \bibinfo {author}
  {\bibfnamefont {Liang}\ \bibnamefont {Jiang}}, \ and\ \bibinfo {author}
  {\bibfnamefont {Hong~X.}\ \bibnamefont {Tang}},\ }\bibfield  {title}
  {{\selectlanguage {english}\enquote {\bibinfo {title} {Magnon dark modes and
  gradient memory},}\ }}\href {\doibase 10.1038/ncomms9914} {\bibfield
  {journal} {\bibinfo  {journal} {Nature Communications}\ }\textbf {\bibinfo
  {volume} {6}},\ \bibinfo {pages} {8914} (\bibinfo {year} {2015})}\BibitemShut
  {NoStop}%
\bibitem [{\citenamefont {Elyasi}\ \emph {et~al.}(2020)\citenamefont {Elyasi},
  \citenamefont {Blanter},\ and\ \citenamefont
  {Bauer}}]{elyasi_resources_2020}%
  \BibitemOpen
  \bibfield  {author} {\bibinfo {author} {\bibfnamefont {Mehrdad}\ \bibnamefont
  {Elyasi}}, \bibinfo {author} {\bibfnamefont {Yaroslav~M.}\ \bibnamefont
  {Blanter}}, \ and\ \bibinfo {author} {\bibfnamefont {Gerrit E.~W.}\
  \bibnamefont {Bauer}},\ }\bibfield  {title} {\enquote {\bibinfo {title}
  {Resources of nonlinear cavity magnonics for quantum information},}\ }\href
  {\doibase 10.1103/PhysRevB.101.054402} {\bibfield  {journal} {\bibinfo
  {journal} {Physical Review B}\ }\textbf {\bibinfo {volume} {101}},\ \bibinfo
  {pages} {054402} (\bibinfo {year} {2020})}\BibitemShut {NoStop}%
\bibitem [{\citenamefont {Harder}\ \emph {et~al.}(2018)\citenamefont {Harder},
  \citenamefont {Yang}, \citenamefont {Yao}, \citenamefont {Yu}, \citenamefont
  {Rao}, \citenamefont {Gui}, \citenamefont {Stamps},\ and\ \citenamefont
  {Hu}}]{harder_level_2018}%
  \BibitemOpen
  \bibfield  {author} {\bibinfo {author} {\bibfnamefont {M.}~\bibnamefont
  {Harder}}, \bibinfo {author} {\bibfnamefont {Y.}~\bibnamefont {Yang}},
  \bibinfo {author} {\bibfnamefont {B.~M.}\ \bibnamefont {Yao}}, \bibinfo
  {author} {\bibfnamefont {C.~H.}\ \bibnamefont {Yu}}, \bibinfo {author}
  {\bibfnamefont {J.~W.}\ \bibnamefont {Rao}}, \bibinfo {author} {\bibfnamefont
  {Y.~S.}\ \bibnamefont {Gui}}, \bibinfo {author} {\bibfnamefont {R.~L.}\
  \bibnamefont {Stamps}}, \ and\ \bibinfo {author} {\bibfnamefont {C.-M.}\
  \bibnamefont {Hu}},\ }\bibfield  {title} {\enquote {\bibinfo {title} {Level
  {Attraction} {Due} to {Dissipative} {Magnon}-{Photon} {Coupling}},}\ }\href
  {\doibase 10.1103/PhysRevLett.121.137203} {\bibfield  {journal} {\bibinfo
  {journal} {Physical Review Letters}\ }\textbf {\bibinfo {volume} {121}},\
  \bibinfo {pages} {137203} (\bibinfo {year} {2018})}\BibitemShut {NoStop}%
\bibitem [{\citenamefont {Grigoryan}\ and\ \citenamefont
  {Xia}(2019)}]{grigoryan_cavity-mediated_2019}%
  \BibitemOpen
  \bibfield  {author} {\bibinfo {author} {\bibfnamefont {Vahram~L.}\
  \bibnamefont {Grigoryan}}\ and\ \bibinfo {author} {\bibfnamefont
  {Ke}~\bibnamefont {Xia}},\ }\bibfield  {title} {\enquote {\bibinfo {title}
  {Cavity-mediated dissipative spin-spin coupling},}\ }\href {\doibase
  10.1103/PhysRevB.100.014415} {\bibfield  {journal} {\bibinfo  {journal}
  {Physical Review B}\ }\textbf {\bibinfo {volume} {100}},\ \bibinfo {pages}
  {014415} (\bibinfo {year} {2019})}\BibitemShut {NoStop}%
\bibitem [{\citenamefont {Yu}\ \emph {et~al.}(2019{\natexlab{a}})\citenamefont
  {Yu}, \citenamefont {Wang}, \citenamefont {Yuan},\ and\ \citenamefont
  {Xiao}}]{yu_prediction_2019}%
  \BibitemOpen
  \bibfield  {author} {\bibinfo {author} {\bibfnamefont {Weichao}\ \bibnamefont
  {Yu}}, \bibinfo {author} {\bibfnamefont {Jiongjie}\ \bibnamefont {Wang}},
  \bibinfo {author} {\bibfnamefont {H.~Y.}\ \bibnamefont {Yuan}}, \ and\
  \bibinfo {author} {\bibfnamefont {Jiang}\ \bibnamefont {Xiao}},\ }\bibfield
  {title} {\enquote {\bibinfo {title} {Prediction of {Attractive} {Level}
  {Crossing} via a {Dissipative} {Mode}},}\ }\href {\doibase
  10.1103/PhysRevLett.123.227201} {\bibfield  {journal} {\bibinfo  {journal}
  {Physical Review Letters}\ }\textbf {\bibinfo {volume} {123}},\ \bibinfo
  {pages} {227201} (\bibinfo {year} {2019}{\natexlab{a}})}\BibitemShut
  {NoStop}%
\bibitem [{\citenamefont {Xu}\ \emph {et~al.}(2019)\citenamefont {Xu},
  \citenamefont {Rao}, \citenamefont {Gui}, \citenamefont {Jin},\ and\
  \citenamefont {Hu}}]{xu_cavity-mediated_2019}%
  \BibitemOpen
  \bibfield  {author} {\bibinfo {author} {\bibfnamefont {Peng-Chao}\
  \bibnamefont {Xu}}, \bibinfo {author} {\bibfnamefont {J.~W.}\ \bibnamefont
  {Rao}}, \bibinfo {author} {\bibfnamefont {Y.~S.}\ \bibnamefont {Gui}},
  \bibinfo {author} {\bibfnamefont {Xiaofeng}\ \bibnamefont {Jin}}, \ and\
  \bibinfo {author} {\bibfnamefont {C.-M.}\ \bibnamefont {Hu}},\ }\bibfield
  {title} {\enquote {\bibinfo {title} {Cavity-mediated dissipative coupling of
  distant magnetic moments: {Theory} and experiment},}\ }\href {\doibase
  10.1103/PhysRevB.100.094415} {\bibfield  {journal} {\bibinfo  {journal}
  {Physical Review B}\ }\textbf {\bibinfo {volume} {100}},\ \bibinfo {pages}
  {094415} (\bibinfo {year} {2019})}\BibitemShut {NoStop}%
\bibitem [{\citenamefont {Yao}\ \emph {et~al.}(2019)\citenamefont {Yao},
  \citenamefont {Yu}, \citenamefont {Zhang}, \citenamefont {Lu}, \citenamefont
  {Gui}, \citenamefont {Hu},\ and\ \citenamefont
  {Blanter}}]{yao_microscopic_2019}%
  \BibitemOpen
  \bibfield  {author} {\bibinfo {author} {\bibfnamefont {Bimu}\ \bibnamefont
  {Yao}}, \bibinfo {author} {\bibfnamefont {Tao}\ \bibnamefont {Yu}}, \bibinfo
  {author} {\bibfnamefont {Xiang}\ \bibnamefont {Zhang}}, \bibinfo {author}
  {\bibfnamefont {Wei}\ \bibnamefont {Lu}}, \bibinfo {author} {\bibfnamefont
  {Yongsheng}\ \bibnamefont {Gui}}, \bibinfo {author} {\bibfnamefont
  {Can-Ming}\ \bibnamefont {Hu}}, \ and\ \bibinfo {author} {\bibfnamefont
  {Yaroslav~M.}\ \bibnamefont {Blanter}},\ }\bibfield  {title} {\enquote
  {\bibinfo {title} {The microscopic origin of magnon-photon level attraction
  by traveling waves: Theory and experiment},}\ }\href {\doibase
  10.1103/PhysRevB.100.214426} {\bibfield  {journal} {\bibinfo  {journal}
  {Phys. Rev. B}\ }\textbf {\bibinfo {volume} {100}},\ \bibinfo {pages}
  {214426} (\bibinfo {year} {2019})}\BibitemShut {NoStop}%
\bibitem [{\citenamefont {Zhang}\ \emph {et~al.}(2019)\citenamefont {Zhang},
  \citenamefont {Wang}, \citenamefont {Hu}, \citenamefont {Shams-Ansari},
  \citenamefont {Ren}, \citenamefont {Fan},\ and\ \citenamefont
  {{Lon\v{c}ar}}}]{zhang_electronically_2019}%
  \BibitemOpen
  \bibfield  {author} {\bibinfo {author} {\bibfnamefont {Mian}\ \bibnamefont
  {Zhang}}, \bibinfo {author} {\bibfnamefont {Cheng}\ \bibnamefont {Wang}},
  \bibinfo {author} {\bibfnamefont {Yaowen}\ \bibnamefont {Hu}}, \bibinfo
  {author} {\bibfnamefont {Amirhassan}\ \bibnamefont {Shams-Ansari}}, \bibinfo
  {author} {\bibfnamefont {Tianhao}\ \bibnamefont {Ren}}, \bibinfo {author}
  {\bibfnamefont {Shanhui}\ \bibnamefont {Fan}}, \ and\ \bibinfo {author}
  {\bibfnamefont {Marko}\ \bibnamefont {{Lon\v{c}ar}}},\ }\bibfield  {title}
  {{\selectlanguage {english}\enquote {\bibinfo {title} {Electronically
  programmable photonic molecule},}\ }}\href {\doibase
  10.1038/s41566-018-0317-y} {\bibfield  {journal} {\bibinfo  {journal} {Nature
  Photonics}\ }\textbf {\bibinfo {volume} {13}},\ \bibinfo {pages} {36}
  (\bibinfo {year} {2019})}\BibitemShut {NoStop}%
\bibitem [{\citenamefont {Baraclough}\ \emph {et~al.}(2019)\citenamefont
  {Baraclough}, \citenamefont {Seetharaman}, \citenamefont {Hooper},\ and\
  \citenamefont {Barnes}}]{baraclough_metamaterial_2019}%
  \BibitemOpen
  \bibfield  {author} {\bibinfo {author} {\bibfnamefont {Milo}\ \bibnamefont
  {Baraclough}}, \bibinfo {author} {\bibfnamefont {Sathya~S.}\ \bibnamefont
  {Seetharaman}}, \bibinfo {author} {\bibfnamefont {Ian~R.}\ \bibnamefont
  {Hooper}}, \ and\ \bibinfo {author} {\bibfnamefont {William~L.}\ \bibnamefont
  {Barnes}},\ }\bibfield  {title} {\enquote {\bibinfo {title} {Metamaterial
  {Analogues} of {Molecular} {Aggregates}},}\ }\href {\doibase
  10.1021/acsphotonics.9b01208} {\bibfield  {journal} {\bibinfo  {journal} {ACS
  Photonics}\ }\textbf {\bibinfo {volume} {6}},\ \bibinfo {pages} {3003--3009}
  (\bibinfo {year} {2019})}\BibitemShut {NoStop}%
\bibitem [{\citenamefont {Prodan}\ \emph {et~al.}(2003)\citenamefont {Prodan},
  \citenamefont {Radloff}, \citenamefont {Halas},\ and\ \citenamefont
  {Nordlander}}]{prodan_hybridization_2003}%
  \BibitemOpen
  \bibfield  {author} {\bibinfo {author} {\bibfnamefont {E.}~\bibnamefont
  {Prodan}}, \bibinfo {author} {\bibfnamefont {C.}~\bibnamefont {Radloff}},
  \bibinfo {author} {\bibfnamefont {N.~J.}\ \bibnamefont {Halas}}, \ and\
  \bibinfo {author} {\bibfnamefont {P.}~\bibnamefont {Nordlander}},\ }\bibfield
   {title} {{\selectlanguage {english}\enquote {\bibinfo {title} {A
  {Hybridization} {Model} for the {Plasmon} {Response} of {Complex}
  {Nanostructures}},}\ }}\href {\doibase 10.1126/science.1089171} {\bibfield
  {journal} {\bibinfo  {journal} {Science}\ }\textbf {\bibinfo {volume}
  {302}},\ \bibinfo {pages} {419--422} (\bibinfo {year} {2003})}\BibitemShut
  {NoStop}%
\bibitem [{\citenamefont {Preston}\ and\ \citenamefont
  {Signorell}(2011)}]{preston_vibron_2011}%
  \BibitemOpen
  \bibfield  {author} {\bibinfo {author} {\bibfnamefont {Thomas~C.}\
  \bibnamefont {Preston}}\ and\ \bibinfo {author} {\bibfnamefont {Ruth}\
  \bibnamefont {Signorell}},\ }\bibfield  {title} {{\selectlanguage
  {english}\enquote {\bibinfo {title} {Vibron and phonon hybridization in
  dielectric nanostructures},}\ }}\href {\doibase 10.1073/pnas.1100170108}
  {\bibfield  {journal} {\bibinfo  {journal} {Proceedings of the National
  Academy of Sciences}\ }\textbf {\bibinfo {volume} {108}},\ \bibinfo {pages}
  {5532--5536} (\bibinfo {year} {2011})}\BibitemShut {NoStop}%
\bibitem [{\citenamefont {Lambert}\ \emph {et~al.}(2016)\citenamefont
  {Lambert}, \citenamefont {Haigh}, \citenamefont {Langenfeld}, \citenamefont
  {Doherty},\ and\ \citenamefont {Ferguson}}]{lambert_cavity-mediated_2016}%
  \BibitemOpen
  \bibfield  {author} {\bibinfo {author} {\bibfnamefont {N.~J.}\ \bibnamefont
  {Lambert}}, \bibinfo {author} {\bibfnamefont {J.~A.}\ \bibnamefont {Haigh}},
  \bibinfo {author} {\bibfnamefont {S.}~\bibnamefont {Langenfeld}}, \bibinfo
  {author} {\bibfnamefont {A.~C.}\ \bibnamefont {Doherty}}, \ and\ \bibinfo
  {author} {\bibfnamefont {A.~J.}\ \bibnamefont {Ferguson}},\ }\bibfield
  {title} {\enquote {\bibinfo {title} {Cavity-mediated coherent coupling of
  magnetic moments},}\ }\href {\doibase 10.1103/PhysRevA.93.021803} {\bibfield
  {journal} {\bibinfo  {journal} {Physical Review A}\ }\textbf {\bibinfo
  {volume} {93}},\ \bibinfo {pages} {021803} (\bibinfo {year}
  {2016})}\BibitemShut {NoStop}%
\bibitem [{\citenamefont {Zare~Rameshti}\ and\ \citenamefont
  {Bauer}(2018)}]{zare_rameshti_indirect_2018}%
  \BibitemOpen
  \bibfield  {author} {\bibinfo {author} {\bibfnamefont {Babak}\ \bibnamefont
  {Zare~Rameshti}}\ and\ \bibinfo {author} {\bibfnamefont {Gerrit E.~W.}\
  \bibnamefont {Bauer}},\ }\bibfield  {title} {\enquote {\bibinfo {title}
  {Indirect coupling of magnons by cavity photons},}\ }\href {\doibase
  10.1103/PhysRevB.97.014419} {\bibfield  {journal} {\bibinfo  {journal}
  {Physical Review B}\ }\textbf {\bibinfo {volume} {97}},\ \bibinfo {pages}
  {014419} (\bibinfo {year} {2018})}\BibitemShut {NoStop}%
\bibitem [{\citenamefont {Junge}\ \emph {et~al.}(2013)\citenamefont {Junge},
  \citenamefont {O'Shea}, \citenamefont {Volz},\ and\ \citenamefont
  {Rauschenbeutel}}]{junge_strong_2013}%
  \BibitemOpen
  \bibfield  {author} {\bibinfo {author} {\bibfnamefont {Christian}\
  \bibnamefont {Junge}}, \bibinfo {author} {\bibfnamefont {Danny}\ \bibnamefont
  {O'Shea}}, \bibinfo {author} {\bibfnamefont {J\"{u}rgen}\ \bibnamefont
  {Volz}}, \ and\ \bibinfo {author} {\bibfnamefont {Arno}\ \bibnamefont
  {Rauschenbeutel}},\ }\bibfield  {title} {\enquote {\bibinfo {title} {Strong
  {Coupling} between {Single} {Atoms} and {Nontransversal} {Photons}},}\ }\href
  {\doibase 10.1103/PhysRevLett.110.213604} {\bibfield  {journal} {\bibinfo
  {journal} {Physical Review Letters}\ }\textbf {\bibinfo {volume} {110}},\
  \bibinfo {pages} {213604} (\bibinfo {year} {2013})},\ \bibinfo {note}
  {publisher: American Physical Society}\BibitemShut {NoStop}%
\bibitem [{\citenamefont {S\"{o}llner}\ \emph {et~al.}(2015)\citenamefont
  {S\"{o}llner}, \citenamefont {Mahmoodian}, \citenamefont {Hansen},
  \citenamefont {Midolo}, \citenamefont {Javadi}, \citenamefont
  {Kir\v{s}ansk\.{e}}, \citenamefont {Pregnolato}, \citenamefont {El-Ella},
  \citenamefont {Lee}, \citenamefont {Song}, \citenamefont {Stobbe},\ and\
  \citenamefont {Lodahl}}]{sollner_deterministic_2015}%
  \BibitemOpen
  \bibfield  {author} {\bibinfo {author} {\bibfnamefont {Immo}\ \bibnamefont
  {S\"{o}llner}}, \bibinfo {author} {\bibfnamefont {Sahand}\ \bibnamefont
  {Mahmoodian}}, \bibinfo {author} {\bibfnamefont {Sofie~Lindskov}\
  \bibnamefont {Hansen}}, \bibinfo {author} {\bibfnamefont {Leonardo}\
  \bibnamefont {Midolo}}, \bibinfo {author} {\bibfnamefont {Alisa}\
  \bibnamefont {Javadi}}, \bibinfo {author} {\bibfnamefont {Gabija}\
  \bibnamefont {Kir\v{s}ansk\.{e}}}, \bibinfo {author} {\bibfnamefont
  {Tommaso}\ \bibnamefont {Pregnolato}}, \bibinfo {author} {\bibfnamefont
  {Haitham}\ \bibnamefont {El-Ella}}, \bibinfo {author} {\bibfnamefont
  {Eun~Hye}\ \bibnamefont {Lee}}, \bibinfo {author} {\bibfnamefont {Jin~Dong}\
  \bibnamefont {Song}}, \bibinfo {author} {\bibfnamefont {S{\o}ren}\
  \bibnamefont {Stobbe}}, \ and\ \bibinfo {author} {\bibfnamefont {Peter}\
  \bibnamefont {Lodahl}},\ }\bibfield  {title} {{\selectlanguage
  {english}\enquote {\bibinfo {title} {Deterministic photon–emitter coupling
  in chiral photonic circuits},}\ }}\href {\doibase 10.1038/nnano.2015.159}
  {\bibfield  {journal} {\bibinfo  {journal} {Nature Nanotechnology}\ }\textbf
  {\bibinfo {volume} {10}},\ \bibinfo {pages} {775--778} (\bibinfo {year}
  {2015})},\ \bibinfo {note} {number: 9 Publisher: Nature Publishing
  Group}\BibitemShut {NoStop}%
\bibitem [{\citenamefont {Lodahl}\ \emph {et~al.}(2017)\citenamefont {Lodahl},
  \citenamefont {Mahmoodian}, \citenamefont {Stobbe}, \citenamefont
  {Rauschenbeutel}, \citenamefont {Schneeweiss}, \citenamefont {Volz},
  \citenamefont {Pichler},\ and\ \citenamefont {Zoller}}]{lodahl_chiral_2017}%
  \BibitemOpen
  \bibfield  {author} {\bibinfo {author} {\bibfnamefont {Peter}\ \bibnamefont
  {Lodahl}}, \bibinfo {author} {\bibfnamefont {Sahand}\ \bibnamefont
  {Mahmoodian}}, \bibinfo {author} {\bibfnamefont {S{\o}ren}\ \bibnamefont
  {Stobbe}}, \bibinfo {author} {\bibfnamefont {Arno}\ \bibnamefont
  {Rauschenbeutel}}, \bibinfo {author} {\bibfnamefont {Philipp}\ \bibnamefont
  {Schneeweiss}}, \bibinfo {author} {\bibfnamefont {J\"{u}rgen}\ \bibnamefont
  {Volz}}, \bibinfo {author} {\bibfnamefont {Hannes}\ \bibnamefont {Pichler}},
  \ and\ \bibinfo {author} {\bibfnamefont {Peter}\ \bibnamefont {Zoller}},\
  }\bibfield  {title} {{\selectlanguage {english}\enquote {\bibinfo {title}
  {Chiral quantum optics},}\ }}\href {\doibase 10.1038/nature21037} {\bibfield
  {journal} {\bibinfo  {journal} {Nature}\ }\textbf {\bibinfo {volume} {541}},\
  \bibinfo {pages} {473--480} (\bibinfo {year} {2017})},\ \bibinfo {note}
  {number: 7638 Publisher: Nature Publishing Group}\BibitemShut {NoStop}%
\bibitem [{\citenamefont {Zhu}\ \emph {et~al.}(2019)\citenamefont {Zhu},
  \citenamefont {Han}, \citenamefont {Zou}, \citenamefont {Xu},\ and\
  \citenamefont {Tang}}]{zhu_magnon-photon_2019}%
  \BibitemOpen
  \bibfield  {author} {\bibinfo {author} {\bibfnamefont {Na}~\bibnamefont
  {Zhu}}, \bibinfo {author} {\bibfnamefont {Xu}~\bibnamefont {Han}}, \bibinfo
  {author} {\bibfnamefont {Chang-Ling}\ \bibnamefont {Zou}}, \bibinfo {author}
  {\bibfnamefont {Mingrui}\ \bibnamefont {Xu}}, \ and\ \bibinfo {author}
  {\bibfnamefont {Hong~X.}\ \bibnamefont {Tang}},\ }\bibfield  {title}
  {\enquote {\bibinfo {title} {Magnon-photon strong coupling for tunable
  microwave circulators},}\ }\href {http://arxiv.org/abs/1912.07128} {\bibfield
   {journal} {\bibinfo  {journal} {arXiv:1912.07128 [physics,
  physics:quant-ph]}\ } (\bibinfo {year} {2019})},\ \bibinfo {note} {arXiv:
  1912.07128}\BibitemShut {NoStop}%
\bibitem [{\citenamefont {Zhang}\ \emph {et~al.}(2020)\citenamefont {Zhang},
  \citenamefont {Galda}, \citenamefont {Han}, \citenamefont {Jin},\ and\
  \citenamefont {Vinokur}}]{zhang_broadband_2020}%
  \BibitemOpen
  \bibfield  {author} {\bibinfo {author} {\bibfnamefont {Xufeng}\ \bibnamefont
  {Zhang}}, \bibinfo {author} {\bibfnamefont {Alexey}\ \bibnamefont {Galda}},
  \bibinfo {author} {\bibfnamefont {Xu}~\bibnamefont {Han}}, \bibinfo {author}
  {\bibfnamefont {Dafei}\ \bibnamefont {Jin}}, \ and\ \bibinfo {author}
  {\bibfnamefont {V.~M.}\ \bibnamefont {Vinokur}},\ }\bibfield  {title}
  {\enquote {\bibinfo {title} {Broadband {Nonreciprocity} {Enabled} by {Strong}
  {Coupling} of {Magnons} and {Microwave} {Photons}},}\ }\href {\doibase
  10.1103/PhysRevApplied.13.044039} {\bibfield  {journal} {\bibinfo  {journal}
  {Physical Review Applied}\ }\textbf {\bibinfo {volume} {13}},\ \bibinfo
  {pages} {044039} (\bibinfo {year} {2020})},\ \bibinfo {note} {publisher:
  American Physical Society}\BibitemShut {NoStop}%
\bibitem [{\citenamefont {Yu}\ \emph {et~al.}(2019{\natexlab{b}})\citenamefont
  {Yu}, \citenamefont {Blanter},\ and\ \citenamefont {Bauer}}]{yu_chiral_2019}%
  \BibitemOpen
  \bibfield  {author} {\bibinfo {author} {\bibfnamefont {Tao}\ \bibnamefont
  {Yu}}, \bibinfo {author} {\bibfnamefont {Yaroslav~M.}\ \bibnamefont
  {Blanter}}, \ and\ \bibinfo {author} {\bibfnamefont {Gerrit E.~W.}\
  \bibnamefont {Bauer}},\ }\bibfield  {title} {\enquote {\bibinfo {title}
  {Chiral {Pumping} of {Spin} {Waves}},}\ }\href {\doibase
  10.1103/PhysRevLett.123.247202} {\bibfield  {journal} {\bibinfo  {journal}
  {Physical Review Letters}\ }\textbf {\bibinfo {volume} {123}},\ \bibinfo
  {pages} {247202} (\bibinfo {year} {2019}{\natexlab{b}})}\BibitemShut
  {NoStop}%
\bibitem [{\citenamefont {Yu}\ \emph {et~al.}(2020{\natexlab{a}})\citenamefont
  {Yu}, \citenamefont {Zhang}, \citenamefont {Sharma}, \citenamefont {Zhang},
  \citenamefont {Blanter},\ and\ \citenamefont {Bauer}}]{yu_magnon_2020}%
  \BibitemOpen
  \bibfield  {author} {\bibinfo {author} {\bibfnamefont {Tao}\ \bibnamefont
  {Yu}}, \bibinfo {author} {\bibfnamefont {Yu-Xiang}\ \bibnamefont {Zhang}},
  \bibinfo {author} {\bibfnamefont {Sanchar}\ \bibnamefont {Sharma}}, \bibinfo
  {author} {\bibfnamefont {Xiang}\ \bibnamefont {Zhang}}, \bibinfo {author}
  {\bibfnamefont {Yaroslav~M.}\ \bibnamefont {Blanter}}, \ and\ \bibinfo
  {author} {\bibfnamefont {Gerrit E.~W.}\ \bibnamefont {Bauer}},\ }\bibfield
  {title} {\enquote {\bibinfo {title} {Magnon {Accumulation} in {Chirally}
  {Coupled} {Magnets}},}\ }\href {\doibase 10.1103/PhysRevLett.124.107202}
  {\bibfield  {journal} {\bibinfo  {journal} {Physical Review Letters}\
  }\textbf {\bibinfo {volume} {124}},\ \bibinfo {pages} {107202} (\bibinfo
  {year} {2020}{\natexlab{a}})},\ \bibinfo {note} {publisher: American Physical
  Society}\BibitemShut {NoStop}%
\bibitem [{\citenamefont {Yu}\ \emph {et~al.}(2020{\natexlab{b}})\citenamefont
  {Yu}, \citenamefont {Zhang}, \citenamefont {Sharma}, \citenamefont
  {Blanter},\ and\ \citenamefont {Bauer}}]{yu_chiral_2020}%
  \BibitemOpen
  \bibfield  {author} {\bibinfo {author} {\bibfnamefont {Tao}\ \bibnamefont
  {Yu}}, \bibinfo {author} {\bibfnamefont {Xiang}\ \bibnamefont {Zhang}},
  \bibinfo {author} {\bibfnamefont {Sanchar}\ \bibnamefont {Sharma}}, \bibinfo
  {author} {\bibfnamefont {Yaroslav~M.}\ \bibnamefont {Blanter}}, \ and\
  \bibinfo {author} {\bibfnamefont {Gerrit E.~W.}\ \bibnamefont {Bauer}},\
  }\bibfield  {title} {\enquote {\bibinfo {title} {Chiral coupling of magnons
  in waveguides},}\ }\href {\doibase 10.1103/PhysRevB.101.094414} {\bibfield
  {journal} {\bibinfo  {journal} {Physical Review B}\ }\textbf {\bibinfo
  {volume} {101}},\ \bibinfo {pages} {094414} (\bibinfo {year}
  {2020}{\natexlab{b}})},\ \bibinfo {note} {publisher: American Physical
  Society}\BibitemShut {NoStop}%
\bibitem [{\citenamefont {Cao}\ \emph {et~al.}(2015)\citenamefont {Cao},
  \citenamefont {Yan}, \citenamefont {Huebl}, \citenamefont {Goennenwein},\
  and\ \citenamefont {Bauer}}]{cao_exchange_2015}%
  \BibitemOpen
  \bibfield  {author} {\bibinfo {author} {\bibfnamefont {Yunshan}\ \bibnamefont
  {Cao}}, \bibinfo {author} {\bibfnamefont {Peng}\ \bibnamefont {Yan}},
  \bibinfo {author} {\bibfnamefont {Hans}\ \bibnamefont {Huebl}}, \bibinfo
  {author} {\bibfnamefont {Sebastian T.~B.}\ \bibnamefont {Goennenwein}}, \
  and\ \bibinfo {author} {\bibfnamefont {Gerrit E.~W.}\ \bibnamefont {Bauer}},\
  }\bibfield  {title} {\enquote {\bibinfo {title} {Exchange magnon-polaritons
  in microwave cavities},}\ }\href {\doibase 10.1103/PhysRevB.91.094423}
  {\bibfield  {journal} {\bibinfo  {journal} {Physical Review B}\ }\textbf
  {\bibinfo {volume} {91}},\ \bibinfo {pages} {094423} (\bibinfo {year}
  {2015})}\BibitemShut {NoStop}%
\bibitem [{\citenamefont {Kittel}(1948)}]{kittel_theory_1948}%
  \BibitemOpen
  \bibfield  {author} {\bibinfo {author} {\bibfnamefont {Charles}\ \bibnamefont
  {Kittel}},\ }\bibfield  {title} {\enquote {\bibinfo {title} {On the {Theory}
  of {Ferromagnetic} {Resonance} {Absorption}},}\ }\href {\doibase
  10.1103/PhysRev.73.155} {\bibfield  {journal} {\bibinfo  {journal} {Physical
  Review}\ }\textbf {\bibinfo {volume} {73}},\ \bibinfo {pages} {155--161}
  (\bibinfo {year} {1948})}\BibitemShut {NoStop}%
\bibitem [{\citenamefont {Jackson}(1998)}]{jackson_classical_1998}%
  \BibitemOpen
  \bibfield  {author} {\bibinfo {author} {\bibfnamefont {John~David}\
  \bibnamefont {Jackson}},\ }\href@noop {} {\emph {\bibinfo {title} {Classical
  Electrodynamics}}}\ (\bibinfo  {publisher} {Wiley, New York},\ \bibinfo
  {year} {1998})\BibitemShut {NoStop}%
\bibitem [{\citenamefont {Sadhana}\ \emph {et~al.}(2009)\citenamefont
  {Sadhana}, \citenamefont {Shinde},\ and\ \citenamefont
  {Murthy}}]{sadhana_synthesis_2009}%
  \BibitemOpen
  \bibfield  {author} {\bibinfo {author} {\bibfnamefont {K.}~\bibnamefont
  {Sadhana}}, \bibinfo {author} {\bibfnamefont {R.~S.}\ \bibnamefont {Shinde}},
  \ and\ \bibinfo {author} {\bibfnamefont {S.~R.}\ \bibnamefont {Murthy}},\
  }\bibfield  {title} {\enquote {\bibinfo {title} {Synthesis of nanocrystalline
  yig using microwave-hydrothermal method},}\ }\href {\doibase
  10.1142/S0217979209063109} {\bibfield  {journal} {\bibinfo  {journal}
  {International Journal of Modern Physics B}\ }\textbf {\bibinfo {volume}
  {23}},\ \bibinfo {pages} {3637--3642} (\bibinfo {year} {2009})},\ \bibinfo
  {note} {publisher: World Scientific Publishing Co.}\BibitemShut {Stop}%
\bibitem [{\citenamefont {Berry}(2009)}]{berry_optical_2009}%
  \BibitemOpen
  \bibfield  {author} {\bibinfo {author} {\bibfnamefont {M.~V.}\ \bibnamefont
  {Berry}},\ }\bibfield  {title} {{\selectlanguage {english}\enquote {\bibinfo
  {title} {Optical currents},}\ }}\href {\doibase
  10.1088/1464-4258/11/9/094001} {\bibfield  {journal} {\bibinfo  {journal}
  {Journal of Optics A: Pure and Applied Optics}\ }\textbf {\bibinfo {volume}
  {11}},\ \bibinfo {pages} {094001} (\bibinfo {year} {2009})}\BibitemShut
  {NoStop}%
\bibitem [{\citenamefont {Aiello}\ \emph {et~al.}(2015)\citenamefont {Aiello},
  \citenamefont {Banzer}, \citenamefont {Neugebauer},\ and\ \citenamefont
  {Leuchs}}]{aiello_transverse_2015}%
  \BibitemOpen
  \bibfield  {author} {\bibinfo {author} {\bibfnamefont {Andrea}\ \bibnamefont
  {Aiello}}, \bibinfo {author} {\bibfnamefont {Peter}\ \bibnamefont {Banzer}},
  \bibinfo {author} {\bibfnamefont {Martin}\ \bibnamefont {Neugebauer}}, \ and\
  \bibinfo {author} {\bibfnamefont {Gerd}\ \bibnamefont {Leuchs}},\ }\bibfield
  {title} {{\selectlanguage {english}\enquote {\bibinfo {title} {From
  transverse angular momentum to photonic wheels},}\ }}\href {\doibase
  10.1038/nphoton.2015.203} {\bibfield  {journal} {\bibinfo  {journal} {Nature
  Photonics}\ }\textbf {\bibinfo {volume} {9}},\ \bibinfo {pages} {789--795}
  (\bibinfo {year} {2015})}\BibitemShut {NoStop}%
\bibitem [{\citenamefont {Aiello}\ and\ \citenamefont
  {Banzer}(2016)}]{aiello_ubiquitous_2016}%
  \BibitemOpen
  \bibfield  {author} {\bibinfo {author} {\bibfnamefont {Andrea}\ \bibnamefont
  {Aiello}}\ and\ \bibinfo {author} {\bibfnamefont {Peter}\ \bibnamefont
  {Banzer}},\ }\bibfield  {title} {{\selectlanguage {english}\enquote {\bibinfo
  {title} {The ubiquitous photonic wheel},}\ }}\href {\doibase
  10.1088/2040-8978/18/8/085605} {\bibfield  {journal} {\bibinfo  {journal}
  {Journal of Optics}\ }\textbf {\bibinfo {volume} {18}},\ \bibinfo {pages}
  {085605} (\bibinfo {year} {2016})}\BibitemShut {NoStop}%
\bibitem [{\citenamefont {Bliokh}\ \emph {et~al.}(2017)\citenamefont {Bliokh},
  \citenamefont {Bekshaev},\ and\ \citenamefont {Nori}}]{bliokh_optical_2017}%
  \BibitemOpen
  \bibfield  {author} {\bibinfo {author} {\bibfnamefont {Konstantin~Y.}\
  \bibnamefont {Bliokh}}, \bibinfo {author} {\bibfnamefont {Aleksandr~Y.}\
  \bibnamefont {Bekshaev}}, \ and\ \bibinfo {author} {\bibfnamefont {Franco}\
  \bibnamefont {Nori}},\ }\bibfield  {title} {\enquote {\bibinfo {title}
  {Optical {Momentum}, {Spin}, and {Angular} {Momentum} in {Dispersive}
  {Media}},}\ }\href {\doibase 10.1103/PhysRevLett.119.073901} {\bibfield
  {journal} {\bibinfo  {journal} {Physical Review Letters}\ }\textbf {\bibinfo
  {volume} {119}},\ \bibinfo {pages} {073901} (\bibinfo {year}
  {2017})}\BibitemShut {NoStop}%
\bibitem [{\citenamefont {Bliokh}\ and\ \citenamefont
  {Nori}(2015)}]{bliokh_transverse_2015}%
  \BibitemOpen
  \bibfield  {author} {\bibinfo {author} {\bibfnamefont {Konstantin~Y.}\
  \bibnamefont {Bliokh}}\ and\ \bibinfo {author} {\bibfnamefont {Franco}\
  \bibnamefont {Nori}},\ }\bibfield  {title} {{\selectlanguage
  {english}\enquote {\bibinfo {title} {Transverse and longitudinal angular
  momenta of light},}\ }}\href {\doibase 10.1016/j.physrep.2015.06.003}
  {\bibfield  {journal} {\bibinfo  {journal} {Physics Reports}\ }\bibinfo
  {series} {Transverse and longitudinal angular momenta of light},\ \textbf
  {\bibinfo {volume} {592}},\ \bibinfo {pages} {1--38} (\bibinfo {year}
  {2015})}\BibitemShut {NoStop}%
\bibitem [{\citenamefont {Banzer}\ \emph {et~al.}(2013)\citenamefont {Banzer},
  \citenamefont {Neugebauer}, \citenamefont {Aiello}, \citenamefont
  {Marquardt}, \citenamefont {Lindlein}, \citenamefont {Bauer},\ and\
  \citenamefont {Leuchs}}]{banzer_photonic_2013}%
  \BibitemOpen
  \bibfield  {author} {\bibinfo {author} {\bibfnamefont {P.}~\bibnamefont
  {Banzer}}, \bibinfo {author} {\bibfnamefont {M.}~\bibnamefont {Neugebauer}},
  \bibinfo {author} {\bibfnamefont {A.}~\bibnamefont {Aiello}}, \bibinfo
  {author} {\bibfnamefont {C.}~\bibnamefont {Marquardt}}, \bibinfo {author}
  {\bibfnamefont {N.}~\bibnamefont {Lindlein}}, \bibinfo {author}
  {\bibfnamefont {T.}~\bibnamefont {Bauer}}, \ and\ \bibinfo {author}
  {\bibfnamefont {G.}~\bibnamefont {Leuchs}},\ }\bibfield  {title}
  {{\selectlanguage {english}\enquote {\bibinfo {title} {The photonic wheel -
  demonstration of a state of light with purely transverse angular momentum},}\
  }}\href {\doibase 10.2971/jeos.2013.13032} {\bibfield  {journal} {\bibinfo
  {journal} {Journal of the European Optical Society - Rapid publications}\
  }\textbf {\bibinfo {volume} {8}} (\bibinfo {year} {2013}),\
  10.2971/jeos.2013.13032}\BibitemShut {NoStop}%
\bibitem [{\citenamefont {Denis}\ \emph {et~al.}(2020)\citenamefont {Denis},
  \citenamefont {Biella}, \citenamefont {Favero},\ and\ \citenamefont
  {Ciuti}}]{denis_permanent_2020}%
  \BibitemOpen
  \bibfield  {author} {\bibinfo {author} {\bibfnamefont {Zakari}\ \bibnamefont
  {Denis}}, \bibinfo {author} {\bibfnamefont {Alberto}\ \bibnamefont {Biella}},
  \bibinfo {author} {\bibfnamefont {Ivan}\ \bibnamefont {Favero}}, \ and\
  \bibinfo {author} {\bibfnamefont {Cristiano}\ \bibnamefont {Ciuti}},\
  }\bibfield  {title} {\enquote {\bibinfo {title} {Permanent {Directional}
  {Heat} {Currents} in {Lattices} of {Optomechanical} {Resonators}},}\ }\href
  {\doibase 10.1103/PhysRevLett.124.083601} {\bibfield  {journal} {\bibinfo
  {journal} {Physical Review Letters}\ }\textbf {\bibinfo {volume} {124}},\
  \bibinfo {pages} {083601} (\bibinfo {year} {2020})},\ \bibinfo {note}
  {publisher: American Physical Society}\BibitemShut {NoStop}%
\bibitem [{\citenamefont {Yang}\ \emph {et~al.}(2019)\citenamefont {Yang},
  \citenamefont {Rao}, \citenamefont {Gui}, \citenamefont {Yao}, \citenamefont
  {Lu},\ and\ \citenamefont {Hu}}]{yang_control_2019}%
  \BibitemOpen
  \bibfield  {author} {\bibinfo {author} {\bibfnamefont {Y.}~\bibnamefont
  {Yang}}, \bibinfo {author} {\bibfnamefont {J.W.}\ \bibnamefont {Rao}},
  \bibinfo {author} {\bibfnamefont {Y.S.}\ \bibnamefont {Gui}}, \bibinfo
  {author} {\bibfnamefont {B.M.}\ \bibnamefont {Yao}}, \bibinfo {author}
  {\bibfnamefont {W.}~\bibnamefont {Lu}}, \ and\ \bibinfo {author}
  {\bibfnamefont {C.-M.}\ \bibnamefont {Hu}},\ }\bibfield  {title} {\enquote
  {\bibinfo {title} {Control of the {Magnon}-{Photon} {Level} {Attraction} in a
  {Planar} {Cavity}},}\ }\href {\doibase 10.1103/PhysRevApplied.11.054023}
  {\bibfield  {journal} {\bibinfo  {journal} {Physical Review Applied}\
  }\textbf {\bibinfo {volume} {11}},\ \bibinfo {pages} {054023} (\bibinfo
  {year} {2019})}\BibitemShut {NoStop}%
\bibitem [{\citenamefont {Li}\ \emph {et~al.}(2019)\citenamefont {Li},
  \citenamefont {Polakovic}, \citenamefont {Wang}, \citenamefont {Xu},
  \citenamefont {Lendinez}, \citenamefont {Zhang}, \citenamefont {Ding},
  \citenamefont {Khaire}, \citenamefont {Saglam}, \citenamefont {Divan},
  \citenamefont {Pearson}, \citenamefont {Kwok}, \citenamefont {Xiao},
  \citenamefont {Novosad}, \citenamefont {Hoffmann},\ and\ \citenamefont
  {Zhang}}]{li_strong_2019}%
  \BibitemOpen
  \bibfield  {author} {\bibinfo {author} {\bibfnamefont {Yi}~\bibnamefont
  {Li}}, \bibinfo {author} {\bibfnamefont {Tomas}\ \bibnamefont {Polakovic}},
  \bibinfo {author} {\bibfnamefont {Yong-Lei}\ \bibnamefont {Wang}}, \bibinfo
  {author} {\bibfnamefont {Jing}\ \bibnamefont {Xu}}, \bibinfo {author}
  {\bibfnamefont {Sergi}\ \bibnamefont {Lendinez}}, \bibinfo {author}
  {\bibfnamefont {Zhizhi}\ \bibnamefont {Zhang}}, \bibinfo {author}
  {\bibfnamefont {Junjia}\ \bibnamefont {Ding}}, \bibinfo {author}
  {\bibfnamefont {Trupti}\ \bibnamefont {Khaire}}, \bibinfo {author}
  {\bibfnamefont {Hilal}\ \bibnamefont {Saglam}}, \bibinfo {author}
  {\bibfnamefont {Ralu}\ \bibnamefont {Divan}}, \bibinfo {author}
  {\bibfnamefont {John}\ \bibnamefont {Pearson}}, \bibinfo {author}
  {\bibfnamefont {Wai-Kwong}\ \bibnamefont {Kwok}}, \bibinfo {author}
  {\bibfnamefont {Zhili}\ \bibnamefont {Xiao}}, \bibinfo {author}
  {\bibfnamefont {Valentine}\ \bibnamefont {Novosad}}, \bibinfo {author}
  {\bibfnamefont {Axel}\ \bibnamefont {Hoffmann}}, \ and\ \bibinfo {author}
  {\bibfnamefont {Wei}\ \bibnamefont {Zhang}},\ }\bibfield  {title} {\enquote
  {\bibinfo {title} {Strong {Coupling} between {Magnons} and {Microwave}
  {Photons} in {On}-{Chip} {Ferromagnet}-{Superconductor} {Thin}-{Film}
  {Devices}},}\ }\href {\doibase 10.1103/PhysRevLett.123.107701} {\bibfield
  {journal} {\bibinfo  {journal} {Physical Review Letters}\ }\textbf {\bibinfo
  {volume} {123}},\ \bibinfo {pages} {107701} (\bibinfo {year}
  {2019})}\BibitemShut {NoStop}%
\bibitem [{\citenamefont {Hou}\ and\ \citenamefont
  {Liu}(2019)}]{hou_strong_2019}%
  \BibitemOpen
  \bibfield  {author} {\bibinfo {author} {\bibfnamefont {Justin~T.}\
  \bibnamefont {Hou}}\ and\ \bibinfo {author} {\bibfnamefont {Luqiao}\
  \bibnamefont {Liu}},\ }\bibfield  {title} {\enquote {\bibinfo {title} {Strong
  {Coupling} between {Microwave} {Photons} and {Nanomagnet} {Magnons}},}\
  }\href {\doibase 10.1103/PhysRevLett.123.107702} {\bibfield  {journal}
  {\bibinfo  {journal} {Physical Review Letters}\ }\textbf {\bibinfo {volume}
  {123}},\ \bibinfo {pages} {107702} (\bibinfo {year} {2019})}\BibitemShut
  {NoStop}%
\bibitem [{\citenamefont {Shao}\ \emph {et~al.}(2018)\citenamefont {Shao},
  \citenamefont {Zhu}, \citenamefont {Chen}, \citenamefont {Zhang},\ and\
  \citenamefont {Yu}}]{shao_spin-orbit_2018}%
  \BibitemOpen
  \bibfield  {author} {\bibinfo {author} {\bibfnamefont {Zengkai}\ \bibnamefont
  {Shao}}, \bibinfo {author} {\bibfnamefont {Jiangbo}\ \bibnamefont {Zhu}},
  \bibinfo {author} {\bibfnamefont {Yujie}\ \bibnamefont {Chen}}, \bibinfo
  {author} {\bibfnamefont {Yanfeng}\ \bibnamefont {Zhang}}, \ and\ \bibinfo
  {author} {\bibfnamefont {Siyuan}\ \bibnamefont {Yu}},\ }\bibfield  {title}
  {{\selectlanguage {english}\enquote {\bibinfo {title} {Spin-orbit interaction
  of light induced by transverse spin angular momentum engineering},}\ }}\href
  {\doibase 10.1038/s41467-018-03237-5} {\bibfield  {journal} {\bibinfo
  {journal} {Nature Communications}\ }\textbf {\bibinfo {volume} {9}},\
  \bibinfo {pages} {1--11} (\bibinfo {year} {2018})}\BibitemShut {NoStop}%
\bibitem [{\citenamefont {Li}(2012)}]{li_radial_2012}%
  \BibitemOpen
  \bibfield  {author} {\bibinfo {author} {\bibfnamefont {Rui}\ \bibnamefont
  {Li}},\ }\href {https://www.osti.gov/biblio/1047870} {\emph {\bibinfo {title}
  {Radial {Eigenmodes} for a {Toroidal} {Waveguide} with {Rectangular} {Cross}
  {Section}}}},\ \bibinfo {type} {Tech. Rep.}\ \bibinfo {number}
  {JLAB-ACP-12-1474; DOE/OR/23177-2264}\ (\bibinfo  {institution} {Thomas
  Jefferson National Accelerator Facility, Newport News, VA (United States)},\
  \bibinfo {year} {2012})\BibitemShut {NoStop}%
\bibitem [{\citenamefont {Stupakov}\ and\ \citenamefont
  {Kotelnikov}(2003)}]{stupakov_shielding_2003}%
  \BibitemOpen
  \bibfield  {author} {\bibinfo {author} {\bibfnamefont {G.~V.}\ \bibnamefont
  {Stupakov}}\ and\ \bibinfo {author} {\bibfnamefont {I.~A.}\ \bibnamefont
  {Kotelnikov}},\ }\bibfield  {title} {\enquote {\bibinfo {title} {Shielding
  and synchrotron radiation in toroidal waveguide},}\ }\href {\doibase
  10.1103/PhysRevSTAB.6.034401} {\bibfield  {journal} {\bibinfo  {journal}
  {Physical Review Special Topics - Accelerators and Beams}\ }\textbf {\bibinfo
  {volume} {6}},\ \bibinfo {pages} {034401} (\bibinfo {year}
  {2003})}\BibitemShut {NoStop}%
\bibitem [{com()}]{comsol}%
  \BibitemOpen
  \href {https://www.comsol.com/} {\enquote {\bibinfo {title} {{COMSOL}
  {Multiphysics}{\circledR} v. 5.4. www.comsol.com. {COMSOL AB}, {Stockholm},
  {Sweden}.}}\ }\BibitemShut {NoStop}%
\bibitem [{\citenamefont {Wang}\ \emph {et~al.}(2016)\citenamefont {Wang},
  \citenamefont {Zhang}, \citenamefont {Zhang}, \citenamefont {Luo},
  \citenamefont {Xiong}, \citenamefont {Wang}, \citenamefont {Li},
  \citenamefont {Hu},\ and\ \citenamefont {You}}]{wang_magnon_2016}%
  \BibitemOpen
  \bibfield  {author} {\bibinfo {author} {\bibfnamefont {Yi-Pu}\ \bibnamefont
  {Wang}}, \bibinfo {author} {\bibfnamefont {Guo-Qiang}\ \bibnamefont {Zhang}},
  \bibinfo {author} {\bibfnamefont {Dengke}\ \bibnamefont {Zhang}}, \bibinfo
  {author} {\bibfnamefont {Xiao-Qing}\ \bibnamefont {Luo}}, \bibinfo {author}
  {\bibfnamefont {Wei}\ \bibnamefont {Xiong}}, \bibinfo {author} {\bibfnamefont
  {Shuai-Peng}\ \bibnamefont {Wang}}, \bibinfo {author} {\bibfnamefont
  {Tie-Fu}\ \bibnamefont {Li}}, \bibinfo {author} {\bibfnamefont {C.-M.}\
  \bibnamefont {Hu}}, \ and\ \bibinfo {author} {\bibfnamefont {J.~Q.}\
  \bibnamefont {You}},\ }\bibfield  {title} {\enquote {\bibinfo {title} {Magnon
  {Kerr} effect in a strongly coupled cavity-magnon system},}\ }\href {\doibase
  10.1103/PhysRevB.94.224410} {\bibfield  {journal} {\bibinfo  {journal}
  {Physical Review B}\ }\textbf {\bibinfo {volume} {94}},\ \bibinfo {pages}
  {224410} (\bibinfo {year} {2016})}\BibitemShut {NoStop}%
\bibitem [{\citenamefont {Papasimakis}\ \emph {et~al.}(2016)\citenamefont
  {Papasimakis}, \citenamefont {Fedotov}, \citenamefont {Savinov},
  \citenamefont {Raybould},\ and\ \citenamefont
  {Zheludev}}]{papasimakis_electromagnetic_2016}%
  \BibitemOpen
  \bibfield  {author} {\bibinfo {author} {\bibfnamefont {N.}~\bibnamefont
  {Papasimakis}}, \bibinfo {author} {\bibfnamefont {V.~A.}\ \bibnamefont
  {Fedotov}}, \bibinfo {author} {\bibfnamefont {V.}~\bibnamefont {Savinov}},
  \bibinfo {author} {\bibfnamefont {T.~A.}\ \bibnamefont {Raybould}}, \ and\
  \bibinfo {author} {\bibfnamefont {N.~I.}\ \bibnamefont {Zheludev}},\
  }\bibfield  {title} {{\selectlanguage {english}\enquote {\bibinfo {title}
  {Electromagnetic toroidal excitations in matter and free space},}\ }}\href
  {\doibase 10.1038/nmat4563} {\bibfield  {journal} {\bibinfo  {journal}
  {Nature Materials}\ }\textbf {\bibinfo {volume} {15}},\ \bibinfo {pages}
  {263--271} (\bibinfo {year} {2016})}\BibitemShut {NoStop}%
\bibitem [{\citenamefont {Talebi}\ \emph {et~al.}(2018)\citenamefont {Talebi},
  \citenamefont {Guo},\ and\ \citenamefont {van}}]{talebi_theory_2018}%
  \BibitemOpen
  \bibfield  {author} {\bibinfo {author} {\bibfnamefont {Nahid}\ \bibnamefont
  {Talebi}}, \bibinfo {author} {\bibfnamefont {Surong}\ \bibnamefont {Guo}}, \
  and\ \bibinfo {author} {\bibfnamefont {Aken Peter~A.}\ \bibnamefont {van}},\
  }\bibfield  {title} {\enquote {\bibinfo {title} {Theory and applications of
  toroidal moments in electrodynamics: their emergence, characteristics, and
  technological relevance},}\ }\href {\doibase 10.1515/nanoph-2017-0017}
  {\bibfield  {journal} {\bibinfo  {journal} {Nanophotonics}\ }\textbf
  {\bibinfo {volume} {7}},\ \bibinfo {pages} {93--110} (\bibinfo {year}
  {2018})}\BibitemShut {NoStop}%
\bibitem [{\citenamefont {Ederer}\ and\ \citenamefont
  {Spaldin}(2007)}]{ederer_towards_2007}%
  \BibitemOpen
  \bibfield  {author} {\bibinfo {author} {\bibfnamefont {Claude}\ \bibnamefont
  {Ederer}}\ and\ \bibinfo {author} {\bibfnamefont {Nicola~A.}\ \bibnamefont
  {Spaldin}},\ }\bibfield  {title} {\enquote {\bibinfo {title} {Towards a
  microscopic theory of toroidal moments in bulk periodic crystals},}\ }\href
  {\doibase 10.1103/PhysRevB.76.214404} {\bibfield  {journal} {\bibinfo
  {journal} {Physical Review B}\ }\textbf {\bibinfo {volume} {76}},\ \bibinfo
  {pages} {214404} (\bibinfo {year} {2007})}\BibitemShut {NoStop}%
\bibitem [{\citenamefont {Heras}(1998)}]{heras_electric_1998}%
  \BibitemOpen
  \bibfield  {author} {\bibinfo {author} {\bibfnamefont {Jos\'{e}A.}\
  \bibnamefont {Heras}},\ }\bibfield  {title} {{\selectlanguage
  {english}\enquote {\bibinfo {title} {Electric and magnetic fields of a
  toroidal dipole in arbitrary motion},}\ }}\href {\doibase
  10.1016/S0375-9601(98)00712-9} {\bibfield  {journal} {\bibinfo  {journal}
  {Physics Letters A}\ }\textbf {\bibinfo {volume} {249}},\ \bibinfo {pages}
  {1--9} (\bibinfo {year} {1998})}\BibitemShut {NoStop}%
\end{thebibliography}
%
\end{document}